\documentclass[twocolumn]{aastex701}

\usepackage[T1]{fontenc} 
\usepackage[utf8]{inputenc} 
\usepackage{amssymb}
\usepackage{float}
\usepackage{placeins}
\usepackage{hyperref}
\usepackage{natbib}
\usepackage{lipsum}

\setcitestyle{notesep={}}

\begin{document}

\title{POSEIDON I: The Dynamical Origins of Transiting Neptunes\footnote{This paper includes data gathered with the 6.5-meter Magellan Telescopes located at Las Campanas Observatory, Chile.}}

\correspondingauthor{Juan I.\ Espinoza-Retamal}
\email{jiespinozar@princeton.edu}

\author[0000-0001-9480-8526]{Juan I.\ Espinoza-Retamal}
\affiliation{Department of Astrophysical Sciences, Princeton University, 4 Ivy Lane, Princeton, NJ 08540, USA}
\email{jiespinozar@princeton.edu}

\author[0000-0002-4265-047X]{Joshua N.\ Winn}
\affiliation{Department of Astrophysical Sciences, Princeton University, 4 Ivy Lane, Princeton, NJ 08540, USA}
\email{jnwinn@princeton.edu}

\author[0000-0002-9158-7315]{Rafael Brahm}
\affil{Facultad de Ingenier\'ia y Ciencias, Universidad Adolfo Ib\'{a}\~{n}ez, Av.\ Diagonal Las Torres 2640, Pe\~{n}alol\'{e}n, Santiago, Chile}
\email{rafael.brahm@uai.cl}

\author[0000-0003-0412-9314]{Cristobal Petrovich}
\affiliation{Department of Astronomy, Indiana University, Bloomington, IN 47405, USA}
\email{cpetrovi@iu.edu}

\author[0000-0001-7409-5688]{Guðmundur Stefánsson} 
\affil{Astrophysics \& Space Institute, Schmidt Sciences, New York, NY 10011, USA}
\affil{Anton Pannekoek Institute for Astronomy, University of Amsterdam, Science Park 904, 1098 XH Amsterdam, The Netherlands}
\email{g.k.stefansson@uva.nl}

\author[0000-0002-5181-0463]{Hareesh Bhaskar} 
\affiliation{Department of Astronomy, Indiana University, Bloomington, IN 47405, USA}
\email{bhareeshg@gmail.com}

\author[0009-0007-0740-0954]{Elise Koo}
\affil{Anton Pannekoek Institute for Astronomy, University of Amsterdam, Science Park 904, 1098 XH Amsterdam, The Netherlands}
\affil{ASTRON, Netherlands Institute for Radio Astronomy, Oude Hoogeveensedijk 4, Dwingeloo 7991 PD, The Netherlands}
\email{e.j.m.koo@uva.nl}

\author[0000-0002-5389-3944]{Andr\'es Jord\'an}
\affil{Facultad de Ingenier\'ia y Ciencias, Universidad Adolfo Ib\'{a}\~{n}ez, Av.\ Diagonal Las Torres 2640, Pe\~{n}alol\'{e}n, Santiago, Chile}
\affil{Departamento de Astronomía, Universidad de Chile, Casilla 36-D, Santiago, Chile}
\affil{El Sauce Observatory --- Obstech, Coquimbo, Chile}
\email{andres.jordan@uai.cl}

\author[0009-0004-8891-4057]{Marcelo Tala Pinto}
\affil{Department of Astronomy, McPherson Laboratory, The Ohio State University, 140 W 18th Ave, Columbus, Ohio 43210, USA}
\email{tala.1@osu.edu}

\author[0000-0002-5945-7975]{Melissa J.\ Hobson}
\affiliation{Observatoire de Genève, Département d'Astronomie, Université de Genève, Chemin Pegasi 51b, 1290 Versoix, Switzerland}
\email{melihobson@gmail.com}

\author[0009-0009-6875-4128]{Hugo Veldhuis}
\affiliation{Anton Pannekoek Institute for Astronomy, University of Amsterdam, Science Park 904, 1098 XH Amsterdam, The Netherlands}
\email{hugo.veldhuis@student.uva.nl}

\author[0000-0003-3047-6272]{Felipe I.\ Rojas} 
\affiliation{Instituto de Astrof\'isica, Pontificia Universidad Cat\'olica de Chile, Av.\ Vicu\~na Mackenna 4860, 7820436 Macul, Santiago, Chile}
\email{firojas@uc.cl}

\author[0009-0008-2801-5040]{Johanna K.\ Teske}
\affiliation{Earth and Planets Laboratory, Carnegie Institution for Science, 5241 Broad Branch Road, NW, Washington, DC 20015, USA}
\affiliation{The Observatories of the Carnegie Institution for Science, 813 Santa Barbara Street, Pasadena, CA 91101, USA}
\email{jteske@carnegiescience.edu}

\author[0000-0003-1305-3761]{R.\ Paul Butler}
\affiliation{Earth and Planets Laboratory, Carnegie Institution for Science, 5241 Broad Branch Road, NW, Washington, DC 20015, USA}
\email{bluaper@gmail.com}

\author[0000-0002-5226-787X]{Jeffrey D.\ Crane}
\affiliation{The Observatories of the Carnegie Institution for Science, 813 Santa Barbara Street, Pasadena, CA 91101, USA}
\email{crane@carnegiescience.edu}

\author[0000-0002-8681-6136]{Stephen Shectman}
\affiliation{The Observatories of the Carnegie Institution for Science, 813 Santa Barbara Street, Pasadena, CA 91101, USA}
\email{shec@carnegiescience.edu}

\author[0000-0003-2527-1475]{Shreyas Vissapragada}
\affiliation{The Observatories of the Carnegie Institution for Science, 813 Santa Barbara Street, Pasadena, CA 91101, USA}
\email{svissapragada@carnegiescience.edu}

\author[0009-0009-2966-7507]{Gavin Boyle}
\affil{El Sauce Observatory --- Obstech, Coquimbo, Chile}
\affil{Cavendish Laboratory, J. J. Thomson Avenue, Cambridge, CB3 0HE, UK}
\email{gavinsboyle@mac.com}

\author[0000-0002-6477-1360]{Rodrigo Leiva}
\affiliation{Instituto de Astrofísica de Andalucía, CSIC, Glorieta de la Astronomía s/n, 18008 Granada, Spain}
\email{rodleiva.astro@gmail.com}

\author[0000-0001-7070-3842]{Vincent Suc}
\affil{Facultad de Ingenier\'ia y Ciencias, Universidad Adolfo Ib\'{a}\~{n}ez, Av.\ Diagonal Las Torres 2640, Pe\~{n}alol\'{e}n, Santiago, Chile}
\affil{El Sauce Observatory --- Obstech, Coquimbo, Chile}
\email{vincent.suc@gmail.com}

\begin{abstract}

We present the first results from the POSEIDON survey, aimed at constraining the dynamical origins of transiting Neptunes through stellar obliquity measurements. We report Rossiter-McLaughlin observations of two Neptunes, TOI-181\,b and TOI-883\,b, obtained with high-resolution spectroscopy from Magellan/PFS and WIYN/NEID. TOI-181\,b is on a 4.5-day orbit with a sky-projected spin-orbit misalignment $\lambda = 32.0_{-6.5}^{+6.3}\,^{\circ}$ and a low eccentricity ($e<0.12$ with $2\sigma$ confidence). TOI-883\,b has a longer orbital period of 10 days with $\lambda = 22_{-14}^{+15}\,^{\circ}$ and eccentricity $e = 0.16 \pm 0.03$. The significant misalignment of TOI-181~b and the significant eccentricity of TOI-883~b are suggestive of high-eccentricity migration for both systems. After adding these and other new measurements to the sample, we analyze the obliquity distribution of the host stars of transiting Neptunes. Earlier studies had suggested that the obliquity distribution is bimodal, with peaks corresponding to aligned orbits and polar orbits; the addition of more measurements has weakened the evidence for bimodality. The current sample appears to be consistent with a population of well-aligned systems and a smaller population with nearly random obliquities. This distribution resembles that observed for more massive planets, suggesting that transiting Jupiters and Neptunes originate from similar dynamical processes.

\end{abstract}


\section{Introduction} 

\begin{figure}
    \centering
    \includegraphics[width=\linewidth]{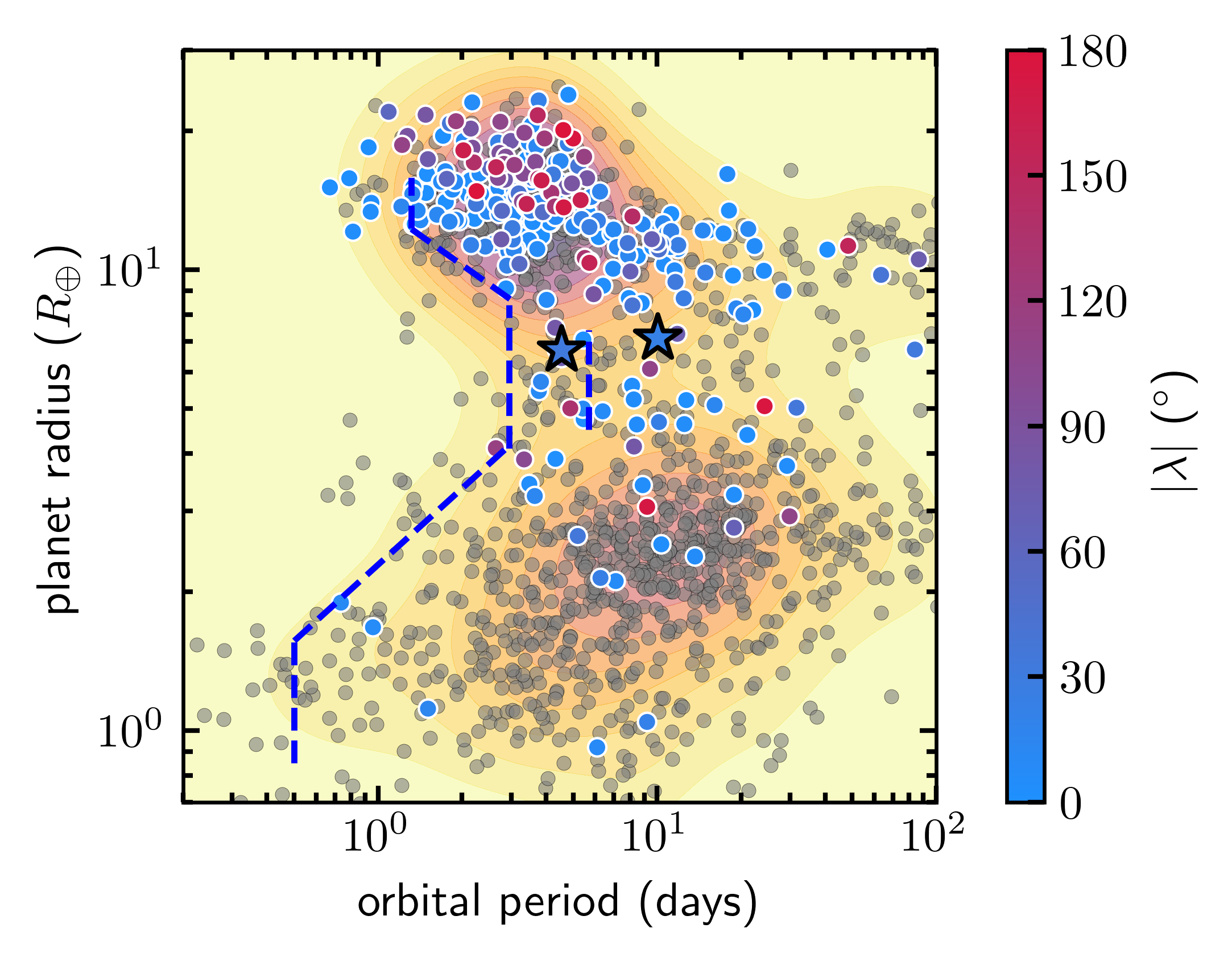}
    \caption{Orbital period versus planet radius for transiting planets. Gray points are from the TEPCat catalog \citep{Southworth2011} as of November 2025. Colored points are those for which the stellar obliquity has been measured, with five-pointed stars highlighting TOI-181~b and TOI-883~b. Dashed blue lines indicate the boundaries of the Neptune desert, ridge, and savanna as defined by \citet{Castro-Gonzalez2024}. The background color conveys the density of data points.}
    \label{fig:neptune_desert}
\end{figure}

Over the last 30 years, more than 6,000 exoplanets have been discovered\footnote{\url{https://exoplanetarchive.ipac.caltech.edu/}}. Most have been discovered using the transit method, with the largest contributions from NASA's Kepler Mission \citep{Borucki2010} and subsequent Transiting Exoplanet Survey Satellite \citep[TESS;][]{Ricker2015}. Most of the known transiting planets orbit within about 0.5~au of their host stars. One of the most prominent features of this population of transiting planets
is the ``hot Neptune desert'' \citep[e.g.,][]{Szabo2011,Mazeh2016}, a dearth of Neptune-sized planets with periods shorter than about 3 days
when compared to smaller planets at similar orbital separations. This feature, evident in the radius versus period diagram displayed in Figure~\ref{fig:neptune_desert}, is possibly explained by the vulnerability of transiting Neptunes to atmospheric loss powered by intense stellar irradiation. \added{For periods beyond about 3 days, the occurrence of Neptune-sized planets increases relative to the hot Neptune desert, in a region} that has
been called the ``Neptune savanna'' \citep{Bourrier2023}. A recent reanalysis of Kepler data by \citet{Castro-Gonzalez2024} identified an apparent peak in the occurrence of transiting Neptunes at 3-6 days separating the desert and the savanna and bearing some resemblance to the ``three-day pile-up'' of hot Jupiters \citep[e.g.,][]{Udry2003,Santerne2016}. This is one of several clues suggesting a possible link between the origins of hot Jupiters and Neptunes \citep[see also][]{Winn2017,Dong2018,Vissapragada2025}.

The origins of these planetary systems can be investigated by studying their architectures, including orbital spacings, companionship statistics, and---the topic of this paper---stellar obliquities. The stellar obliquity, $\psi$, is the angle between the stellar spin vector and the planet's orbital angular momentum vector \citep[see, e.g.,][]{Albrecht2022}. Hundreds of obliquity measurements have been obtained through observations of the Rossiter-McLaughlin (RM) effect \citep{Rossiter1924,McLaughlin1924}, which is sensitive to the sky-projected obliquity, $\lambda$. Recent studies have suggested that stars with transiting planets have a bimodal obliquity distribution, with one population of well-aligned systems ($\psi\sim0^{\circ}$) and another of nearly polar ($\psi\sim90^{\circ}$) systems \citep[e.g.,][]{Albrecht2021,Bourrier2023,Attia2023}. Such a distribution would be an important clue about planet migration models
and possible tidal interactions. However, more recent analyses have questioned the statistical significance of the bimodal obliquity distribution \citep[e.g.,][]{Siegel2023,Dong2023,Rossi2025}.

Although obliquities offer a useful probe of formation pathways, current measurements are mainly limited to gas giant planets (see the colored points in Figure \ref{fig:neptune_desert}). Since the amplitude of the RM effect scales with the amount of starlight blocked by the planet during the transit \citep[e.g.,][]{Triaud2018}, the large sizes of gas giants make RM observations easier, explaining why transiting Jupiters dominate the existing sample. This is unfortunate because Neptunes are an order of magnitude more common than Jupiters within 0.5 au of Sun-like stars \citep[e.g.,][]{Howard2010}. A sample biased toward giant planets may therefore not reflect the dynamical processes that govern the broader, more diverse population of transiting planets.

Despite the limited number of available measurements, recent studies have suggested that the preference for polar orbits may be particularly strong among transiting Neptunes \citep{Espinoza-Retamal2024, Knudstrup2024, Handley2025, Rossi2025}. If confirmed, this would imply distinct dynamical histories for transiting Jupiters and Neptunes. However, the current sample is too small to draw firm conclusions, motivating an obliquity survey targeting smaller planets.

This paper is the first in what is intended to be a series of obliquity-based studies of the dynamical origins of transiting Neptunes. Here, we present RM observations and obliquity measurements for two Neptune systems: TOI-181 \citep{Mistry2023}, which hosts a Neptune in the ridge, and TOI-883 \citep{Rojas2025}, which hosts a Neptune in the savanna (see Figure~\ref{fig:neptune_desert}). 

We have structured this paper as follows. Section~\ref{sec:observations} describes new and archival observations of both systems. Section~\ref{sec:analysis} presents our analysis methods. Section~\ref{sec:discussion} describes the results and discusses their implications, both for each individual system and for the entire population of transiting Neptunes. Finally, Section~\ref{sec:conclusion} summarizes our findings and conclusions.

\section{Observations}\label{sec:observations}

\subsection{PFS Transit Spectroscopy}

\begin{figure*}[t!]
    \digitalasset
    \centering
    \includegraphics[width=\linewidth]{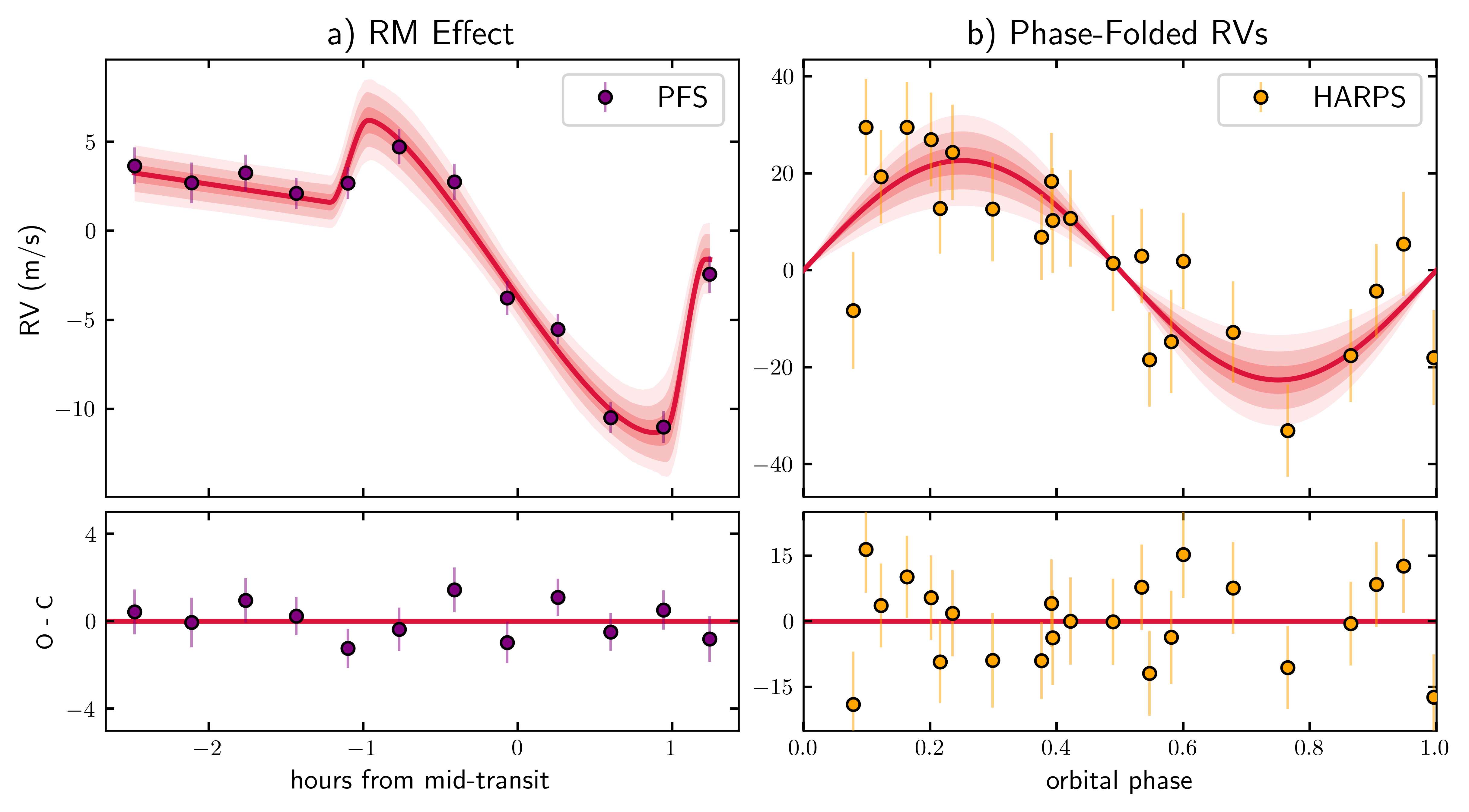}
    \includegraphics[width=0.7\linewidth]{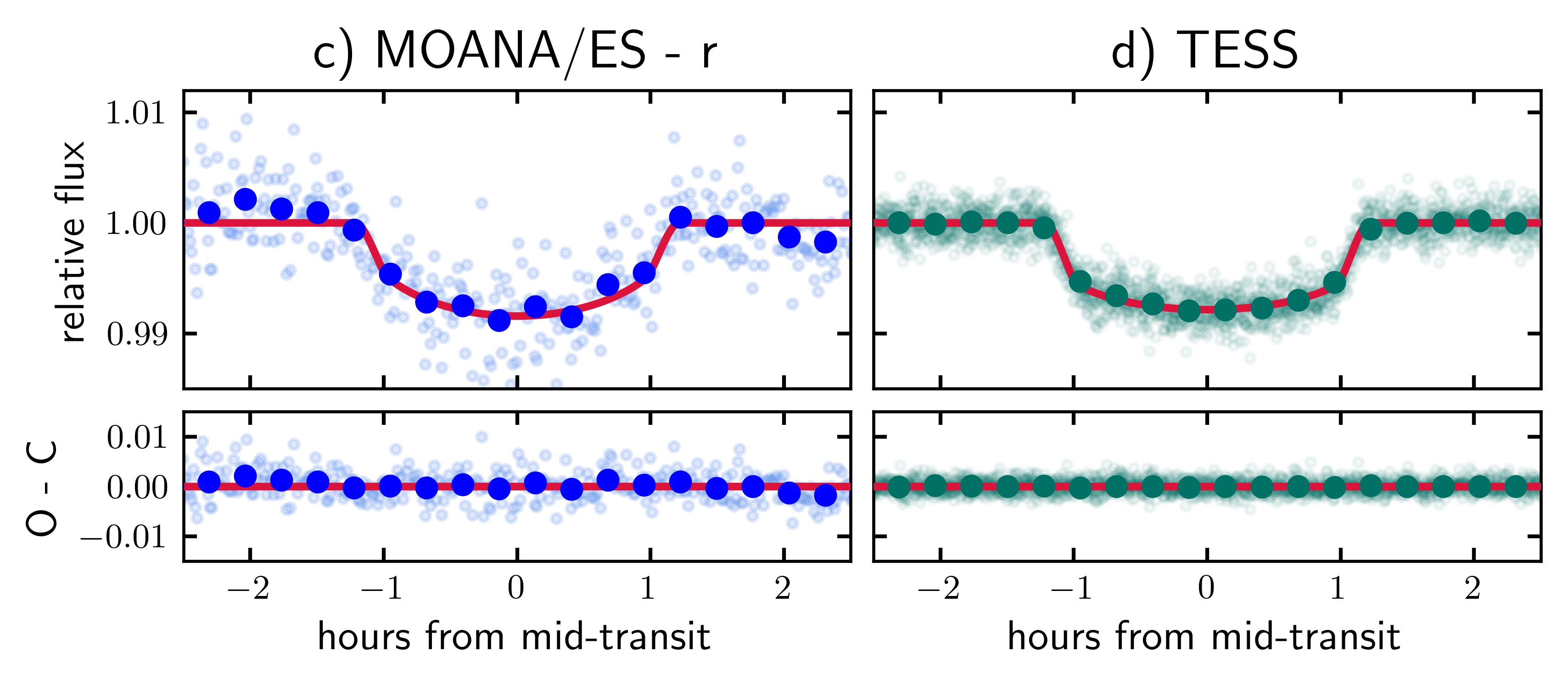}
    \caption{Radial-velocity and photometric observations of TOI-181. In all cases, the red curves are best-fit models and the residuals
    are plotted beneath the data. The error bars include a white noise jitter term added in quadrature. a) PFS velocities spanning a transit and exhibiting the RM effect. b) HARPS velocities across all orbital phases. c) MOANA photometry of the same transit observed with PFS. d) Phase-folded photometry based on TESS observations
    with 2-minute cadence (green). In panels c and d,  the darker points are time-averaged data. Data behind the figure are available electronically.}
    \label{fig:fit_181}
\end{figure*}

We observed one transit of TOI-181 b on 2025 July 3 between 05:01 and 08:44~UTC, using Carnegie's Planet Finder Spectrograph \citep[PFS;][]{Crane2006,Crane2008,Crane2010} mounted on the Magellan Clay 6.5\,m Telescope at Las Campanas Observatory, Chile. PFS covers the wavelength range 391--734~nm at a default resolving power of $R\approx130,\!000$ ($1\times2$ binning). Due to the faintness of TOI-181 ($V=11.2$ mag), for these observations we used $3\times3$ binning, resulting in a lower resolution of $R\approx110,\!000$. We obtained 12 spectra of the host star with an exposure time of 1200\,s. In order to derive precise radial velocities (RVs), we also constructed
an iodine-free template spectrum based on
three consecutive 1200\,s PFS exposures with the same $3\times3$ binning on 2025~July 4. The spectra were processed and the RVs were extracted using a customized \texttt{IDL} pipeline described by \citet{Butler1996}. Each processed
spectrum of TOI-181 has a median signal-to-noise ratio of 53 per pixel at 550~nm, resulting in a median RV uncertainty of $0.97~{\rm m~s^{-1}}$ based on the scatter seen in RVs estimated from different chunks of the spectrum. The RVs of TOI-181 are shown in Figure~\ref{fig:fit_181}, and are available as the data behind the figure.

\subsection{NEID Transit Spectroscopy}

\begin{figure*}[t!]
    \digitalasset
    \centering
    \includegraphics[width=\linewidth]{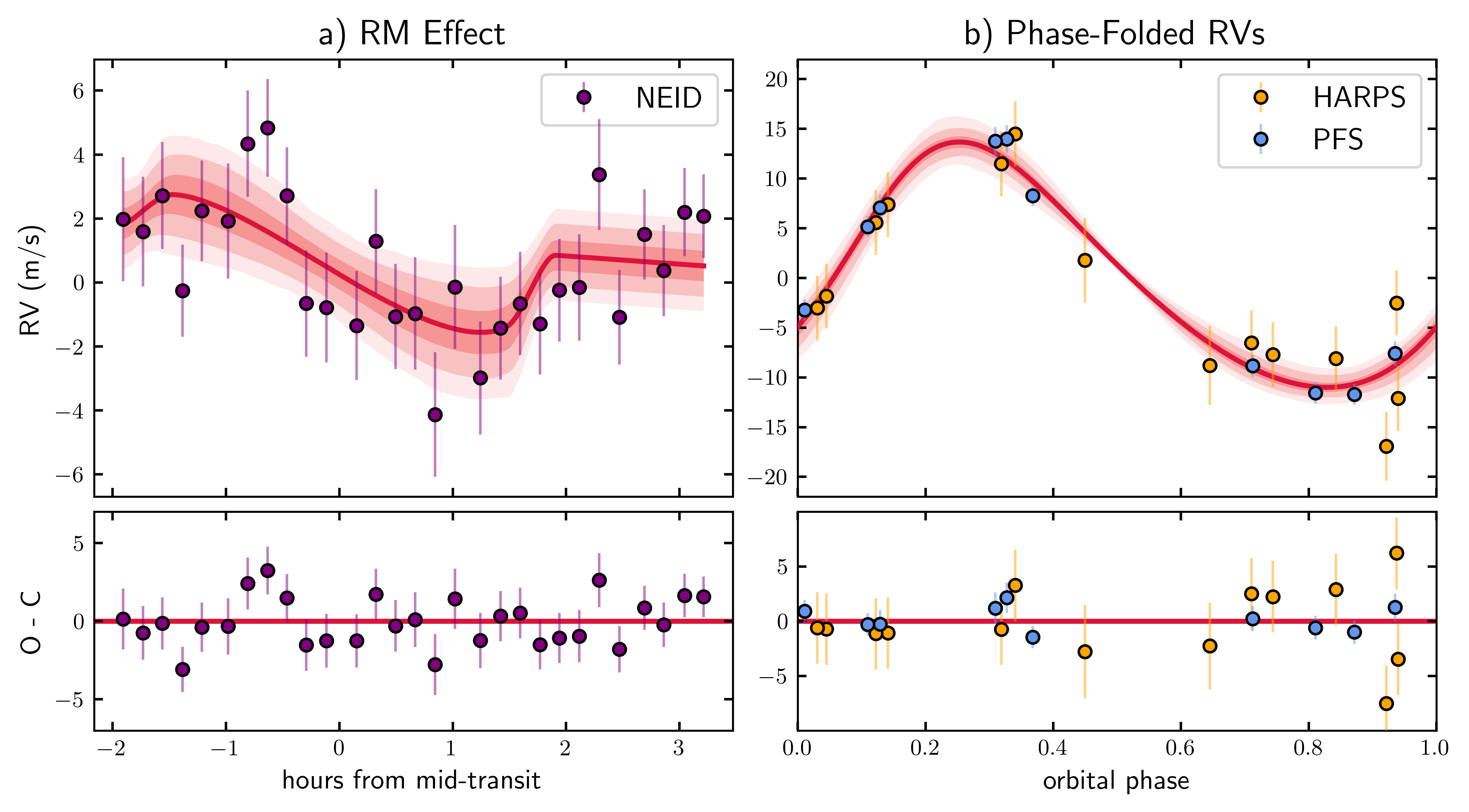}
    \includegraphics[width=0.7\linewidth]{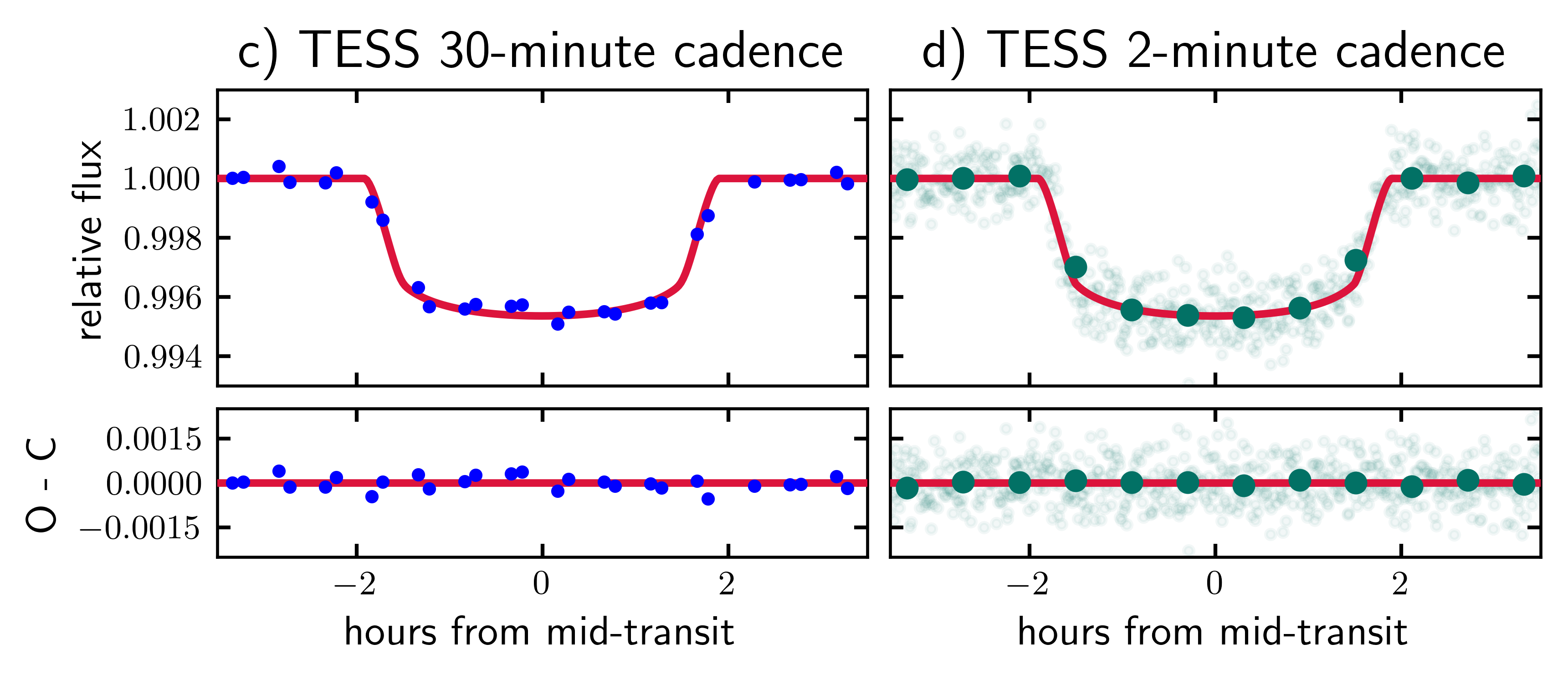}
    \caption{Radial-velocity and photometric observations of TOI-883. In all cases, the red curves are best-fit models and the residuals are plotted beneath the data. The RV error bars include a white noise jitter term added in quadrature. a) NEID velocities spanning a transit and exhibiting the RM effect. b) HARPS and PFS velocities across all orbital phases. c) Phase-folded photometry based on TESS observations
    with 30-minute cadence (blue). d) Phase-folded photometry based on TESS observations
    with 2-minute cadence (green). The darker points are time-averaged data. Data behind the figure are available electronically.}
    \label{fig:fit_883}
\end{figure*}

We observed one transit of TOI-883 b using the NEID spectrograph \citep{Schwab2016}, installed on the WIYN 3.5\,m telescope at Kitt Peak Observatory in Arizona. NEID is an environmentally stabilized, fiber-fed echelle spectrograph covering the wavelength range 380--930~nm with a resolving power of $R\approx110,\!000$ \citep{Stefansson2016,Robertson2019,Kanodia2018,Halverson2016}. The transit was observed on 1~November~2025 between 07:40 and 12:47~UTC. We obtained 29 spectra with an exposure time of 600\,s. The spectra were processed with the NEID data reduction pipeline \citep[version 1.4.2; e.g.,][]{Bender2022}, which also delivered RVs using the cross-correlation function (CCF) method. The final NEID spectra have a median signal-to-noise ratio of 41 per 1D extracted pixel at 550~nm, resulting in a median RV uncertainty of 2.3 m~s$^{-1}$, based on the scatter observed among the weighted order-by-order CCF velocities. As a complement to the CCF-based RVs, we extracted RVs using \texttt{NEID-SERVAL}, a spectral-matching code based on the SpEctrum Radial Velocity AnaLyzer \citep[\texttt{SERVAL};][]{Zechmeister2018} code and adapted for NEID by \citet{Stefansson2022}. The \texttt{NEID-SERVAL} RVs have a median RV uncertainty of 1.6 m s$^{-1}$, based on the scatter observed among the weighted order-by-order velocities. We adopted the \texttt{NEID-SERVAL} RVs for the analysis because they are more precise, but consistent results are obtained with the CCF RVs. The \texttt{NEID-SERVAL} RVs for TOI-883, along with the best model, are shown in Figure \ref{fig:fit_883} and are available as the data behind the figure.

\subsection{Observatoire Moana Photometry}

Simultaneously with the PFS observations, we observed the transit of TOI-181 b using the station of the Observatoire Moana located in El Sauce Observatory in Chile. Observatoire Moana is a global network of small-aperture robotic optical telescopes \citep[see, e.g.,][]{Trifonov2023,Brahm2023,Brahm2025}. The El Sauce station consists of a 0.6~m robotic telescope coupled to an Andor iKon-L~936 deep depletion 2k $\times$ 2k CCD with a scale of 0.67$^{\prime\prime}$ per pixel. We observed the transit through a Sloan $r$ filter. A custom data reduction pipeline automatically performed the CCD reductions and aperture photometry for the brightest stars in the field. The pipeline also generated the differential light curve of the target star by identifying the optimal comparison stars based on color, brightness, and proximity to the target \citep[see, e.g.,][]{Jordan2019,Brahm2020}. The final light curve, along with the best model, is shown in Figure \ref{fig:fit_181}.

\subsection{TESS Photometry}

Since the initial reports of both planets \citep{Mistry2023,Rojas2025}, TESS has obtained additional
data. We took advantage of these new TESS data to obtain more precise ephemerides and jointly fit the light curves with the spectroscopic observations. We retrieved the TESS light curves from the Mikulski Archive for Space Telescopes using the \texttt{lightkurve} code \citep{lightkurve}. For TOI-181, we used the 2-minute cadence light curves from Sectors 2, 29, and 69 \added{\citep{TESS_2min}}. For TOI-883, we used the 30-minute cadence light curve from Sector 6 \added{\citep{TESS_30min}} and the 2-minute light curves from Sectors 33 and 87 \added{\citep{TESS_2min}}. All these data were processed with the TESS Science Processing Operations Center pipeline \citep{spoc}, and as such, were corrected for pointing and focus-related instrumental signatures, discontinuities resulting from radiation events in the CCD detectors, outliers, and contributions to the recorded flux from nearby stars. TESS transit observations of TOI-181 and TOI-883, along with the best-fitting models, are shown in Figures \ref{fig:fit_181} and \ref{fig:fit_883}, respectively.

\subsection{Archival Spectroscopy}

To precisely constrain the parameters of the planets and their orbits, we used archival spectroscopic observations of TOI-181 and TOI-883. For TOI-181, we included the 23 HARPS \citep{Mayor2003} RVs presented by \citet{Mistry2023}. For TOI-883, we included the 14 HARPS and 10 PFS RVs presented by \citet{Rojas2025}. These archival observations of TOI-181 and TOI-883, along with the best-fit models, are shown in Figures \ref{fig:fit_181} and \ref{fig:fit_883}.

\section{Analysis}\label{sec:analysis}

\subsection{Stellar Parameters}\label{sec:stellar}

The POSEIDON team intends to perform a homogeneous analysis of the survey targets, including uniformly derived stellar parameters. We have adopted the method described by \citet{Brahm2019}. The stellar parameters for TOI-883 reported by \citet{Rojas2025} were already estimated using the same methods, and needed no modification.
For TOI-181, we applied the method of
\citet{Brahm2019} to derive
stellar parameters based on the out-of-transit HARPS spectra obtained by \citet{Mistry2023}. Briefly, the method consists of two steps that are repeated iteratively. In the first step, we compute the stellar atmospheric parameters using the \texttt{zaspe} package \citep{zaspe}, which
compares the co-added high-resolution spectrum with a grid of synthetic spectra to identify the closest match. The search is performed in the spectral regions that are most sensitive to changes in the stellar parameters, and error bars are computed through Monte Carlo simulations. In the second step, we compute the stellar physical parameters by fitting 
stellar-evolutionary models to the observed spectral energy distribution. We fit the available broadband apparent magnitudes and the parallax from Gaia Data Release 3 \citep[DR3;][]{GaiaDR3}
to synthetic magnitudes generated from the \texttt{PARSEC} isochrones \citep{parsec} and the \citet{Cardelli1989}
model for interstellar extinction. In this step, the stellar temperature derived with \texttt{zaspe} is used as a prior, while the metallicity is held fixed. From the stellar mass and radius obtained with the second step, we calculate a more precise $\log{g}$, which is held fixed in a new iteration of the first step. We iterated between the two procedures until reaching convergence, which happens when two consecutive \texttt{zaspe} runs deliver the same values of $T_{\rm eff}$ and [Fe/H]. As these parameters come from a grid search, for small changes in $\log{g}$ ($\Delta \log{g} < 0.01$), the model with the best matching $T_{\rm eff}$ and [Fe/H] will be the same. The derived stellar parameters are within $2\sigma$ of those reported in the literature and are presented in Table~\ref{tab:stellar}.

\begin{deluxetable*}{llccr}
\tablecaption{Stellar properties of TOI-181 and TOI-883.\label{tab:stellar}}
\tablecolumns{4}
\tablewidth{0pt}
\tablehead{Parameter & Description & TOI-181 & TOI-883 & Reference}
\startdata
RA & Right Ascension (J2015.5) & 23h28m41.41s & 06h53m28.57s & \citet{GaiaDR3}\\
Dec & Declination (J2015.5) & $-$34d29m29.08s & $-$06d30m28.21s & \citet{GaiaDR3}\\
pm$^{\rm RA}$ & Proper motion in RA (mas yr$^{-1}$) & 111.503$\pm$0.015 & $-$50.093$\pm$0.021 & \citet{GaiaDR3}\\
pm$^{\rm Dec}$ & Proper motion in DEC (mas yr$^{-1}$) & $-$2.642$\pm$0.015 & $-$76.006$\pm$0.019 & \citet{GaiaDR3}\\
$\pi$ & Parallax (mas) & 10.384$\pm$0.018 & 9.759$\pm$0.017 & \citet{GaiaDR3} \\
$d$ & Distance (pc) & 96.3$\pm$0.2& 102.5$\pm$0.2 & \citet{GaiaDR3} \\
\hline
T & TESS magnitude (mag) & 10.465$\pm$0.006 & 9.372$\pm$0.006 & \citet{Stassun2018,Stassun2019}\\
B  & B-band magnitude (mag) & 12.235$\pm$0.211 & 10.57$\pm$0.07 & \citet{apass}\\
V  & V-band magnitude (mag) & 11.191$\pm$0.019 & 9.957$\pm$0.005 & \citet{apass}\\
G  & Gaia G-band magnitude (mag) & 11.0863$\pm$0.0006 & 9.8326$\pm$0.0002 & \citet{GaiaDR3}\\
G$_{\rm BP}$ & Gaia BP-band magnitude (mag) & 11.6162$\pm$0.0019 & 10.1792$\pm$0.0003 & \citet{GaiaDR3}\\
G$_{\rm RP}$ & Gaia RP-band magnitude (mag) & 10.4046$\pm$0.0014 & 	9.3175$\pm$0.0002 &  \citet{GaiaDR3}\\
J & 2MASS J-band magnitude (mag) & 9.627$\pm$0.029 & 8.749$\pm$0.026 & \citet{2mass}\\
H & 2MASS H-band magnitude (mag) & 9.140$\pm$0.022 & 8.437$\pm$0.038 & \citet{2mass}\\
K$_s$ & 2MASS K$_s$-band magnitude (mag) & 9.047$\pm$0.021 & 8.37$\pm$0.023 & \citet{2mass}\\
\hline
$T_{\rm eff}$ & Effective temperature (K) & 4820$\pm$80 & 5697$\pm$80 & This work\\
$\log{g}$ & Surface gravity (with $g$ in cm\,s$^{-2}$) &  4.59$\pm$0.02 & $4.43\pm0.02$ & This work\\
$[$Fe/H$]$ & Metallicity (dex) & +0.14$\pm$0.05 & $+0.02\pm0.04$ & This work\\
$v\sin{i_\star}$ & Projected rotational velocity (km s$^{-1}$) & 1.9$\pm$0.7 & 2.5$\pm$0.3 & This work\\
$M_{\star}$ & Mass ($M_\odot$) & 0.802$\pm$0.021 & $0.96\pm0.05$ & This work\\
$R_{\star}$ & Radius ($R_\odot$) & 0.750$\pm0.007$ & $0.99\pm0.01$ & This work\\
$L_{\star}$ & Luminosity ($L_\odot$) & 0.286$_{-0.006}^{+0.010}$ & $0.93^{+0.04}_{-0.03}$ & This work\\
$A_{V}$ & Visual extinction (mag) & 0.055$_{-0.038}^{+0.058}$ & $0.08^{+0.06}_{-0.05}$ & This work\\
Age & Age (Gyr) & $5.5_{-3.3}^{+3.8}$ & $6.5^{+2.0}_{-2.1}$ & This work\\
$\rho_\star$ & Bulk density (g cm$^{-3}$) & 2.69$\pm$0.13 & $1.40\pm0.15$ & This work\\
\enddata
\tablecomments{The stellar parameters of TOI-883 come from \citet{Rojas2025}, who determined them
using the same procedure described in Section \ref{sec:stellar}. The uncertainties do not take into account possible systematic differences among different stellar evolutionary models \citep{Tayar2022}. The TESS magnitude is shown only for reference and was not included in our stellar analysis.}
\end{deluxetable*}

\subsection{Photometry}\label{sec:phot}

In order to refine the orbital ephemeris of TOI-181~b and TOI-883~b, and to look for transit timing variations (TTVs), we performed an analysis of all the photometric observations described in Section \ref{sec:observations} with the \texttt{juliet} code \citep{juliet}, which uses \texttt{batman} \citep{batman} for the transit model and  \texttt{dynesty} \citep{dynesty2} for sampling
the posterior probability distributions. For each target, we placed uniform priors on the impact parameter $b$ and radius ratio $R_p/R_{\star}$, with an informative Gaussian prior on the stellar density $\rho_\star$ based on estimates of the
stellar mass and radius described in Section \ref{sec:stellar}. We assumed the limb-darkening law to be quadratic and sampled the coefficients
$q_1$ and $q_2$ defined by \citet{Kipping13} with uniform priors. We placed broad Gaussian priors (with a width
of 1~day) on the time
of each transit midpoint, based on the expected values calculated from the orbital period and time of mid-transit from the discovery paper for each planet.

To account for variability and systematic noise in the TESS light curves, we included a Matern-3/2 Gaussian Process (GP) as implemented in \texttt{celerite} \citep{celerite} and available in \texttt{juliet}. For TOI-883, we considered the 30-minute and 2-minute TESS light curves as coming from two different instruments, each with its own GP kernel to account for differences in variability captured in different epochs and cadences, while sharing the limb darkening coefficients. From this analysis, we ruled out the presence of TTVs larger than $\sim5$ minutes for TOI-181~b and $\sim2$ minutes for TOI-883~b. Additionally, we obtained improved orbital ephemerides for both planets, and constructed a detrended TESS light curve for each system that was used in the global modeling described below. 

\subsection{Global Modeling}

To constrain the stellar obliquity in the two systems studied here, we performed a joint analysis of all the photometric and spectroscopic observations presented in Section \ref{sec:observations} using the \texttt{ironman}\footnote{\url{https://github.com/jiespinozar/ironman}} package \citep{Espinoza-Retamal2023b,Espinoza-Retamal2024}, which jointly fits transit RVs and photometry. To model the RM effect, \texttt{ironman} uses the \texttt{rmfit} code \citep{Stefansson2022}, which is based on the equations from \citet{Hirano2010}. To model the transit light curves, \texttt{ironman} uses \texttt{batman} \citep{batman}. To model the out-of-transit RVs, \texttt{ironman} uses \texttt{radvel} \citep{Fulton18}. Finally, to sample the posteriors, the code uses the \texttt{dynesty} dynamic nested sampler \citep{dynesty2}. \added{In both cases, we adopted 4500 live points, and the default configurations of the sampler for the bounding and sampling methods (i.e., multi-ellipsoidal bounding and random-walk sampling). The runs were stopped when the change in the logarithm of the Bayesian evidence ($\log{Z}$) was less than 0.01, which is the default convergence criterion.} Since the stellar rotation periods are not known, the stellar inclination angle is not well constrained. As a result, the sky-projected stellar obliquity can be determined from the data but the true obliquity is undetermined.

In this analysis, we only considered the TESS data within 10 hours of the transit midpoint to reduce the computational cost. We included independent jitter terms for each instrument to account for possible instrumental systematics. In both systems, we placed uniform priors on almost all the parameters. The exceptions were the stellar densities, for which we placed informative Gaussian priors based on the values derived in Section \ref{sec:stellar}. We also placed Gaussian priors on $\beta$, which accounts for the intrinsic linewidth due to instrumental and macroturbulence broadening \citep[see][]{Hirano2010}. We considered an instrumental broadening of 2.7 km s$^{-1}$ because of the PFS and NEID resolutions ($R\approx110,\!000$), and for the macroturbulence broadening, we used the macroturbulence law from \citet{Valenti2005} to estimate it using the stellar effective temperatures reported in Table \ref{tab:stellar}. We added the instrumental and macroturbulence broadening in quadrature to set our priors for each target, with an uncertainty of 2 km s$^{-1}$. All priors and results obtained from the posterior distributions for each system are shown in Tables \ref{tab:fit_181} and \ref{tab:fit_883}.

\begin{deluxetable*}{llcr}[h!]
\tablecaption{Summary of priors and posteriors of the \texttt{ironman} fit for TOI-181. \label{tab:fit_181}}
\tablewidth{70pt}
\tablehead{Parameter & Description & Prior & Posterior}
\startdata
$\lambda$ & Sky-projected stellar obliquity ($^{\circ}$) & $\mathcal{U}(-180,180)$ & $32.0_{-6.5}^{+6.3}$ \\
$v\sin{i_\star}$ & Projected rotational velocity (km s$^{-1}$) & $\mathcal{U}(0,10)$ & $2.1\pm0.2$ \\
$\rho_\star$ & Stellar density (g cm$^{-3}$) & $\mathcal{G}(2.69,0.13)$ & $2.64\pm0.13$ \\
\hline
$P$ & Orbital period (days) & $\mathcal{U}(4.532045,4.532061)$ & $4.532055\pm0.000001$ \\ 
$t_0$ & Transit midpoint (BJD) & $\mathcal{U}(2458353.5844,2458353.5885)$ & $2458353.5861\pm0.0002$ \\
$b$ & Impact parameter & $\mathcal{U}(0,1)$ & $0.44\pm0.04$ \\
$i$ & Orbital inclination ($^{\circ}$) & \nodata & $88.2\pm0.2$ \\
$R_p/R_\star$ & Radius ratio & $\mathcal{U}(0,1)$ & $0.082\pm0.001$ \\
$K$ & RV semiamplitude \added{(m s$^{-1}$)} & $\mathcal{U}(0,1000)$ & $22\pm3$ \\
$e$ & Eccentricity & \nodata & 0 (fixed, $<0.12$ at $2\sigma$)\\
$a/R_\star$ & Scaled semimajor axis & \nodata & $14.2\pm0.2$ \\
$R_p$ & Planet radius ($R_\oplus$)& \nodata & $6.7\pm0.2$ \\
$a$ & Semimajor axis (au)& \nodata & $0.049\pm0.002$ \\
$M_p$ & Planet mass ($M_\oplus$)& \nodata & $49.8^{+6.7}_{-6.5}$ \\
$\rho_p$ & Planet density (g cm$^{-3}$)& \nodata & $0.91\pm0.15$ \\
\hline
$q_1^{\rm PFS}$ & PFS linear limb darkening parameter & $\mathcal{U}(0,1)$ & $0.33_{-0.21}^{+0.33}$ \\
$q_2^{\rm PFS}$ & PFS quadratic limb darkening parameter & $\mathcal{U}(0,1)$ & $0.32_{-0.23}^{+0.38}$ \\
$\gamma_{\rm PFS}$ & PFS RV offset (m s$^{-1}$)& $\mathcal{U}(-100,100)$ & $-3\pm1$ \\
$\sigma_{\rm PFS}$ & PFS RV jitter (m s$^{-1}$)& $\mathcal{LU}(10^{-3},100)$ & $0.04_{-0.03}^{+0.40}$ \\
$\beta$ & Intrinsic stellar line width (km s$^{-1}$) & $\mathcal{G}(3.9,2.0)$ & $4.8_{-1.5}^{+1.7}$ \\
$\gamma_{\rm HARPS}$ & HARPS RV offset (m s$^{-1}$)& $\mathcal{U}(-4900,-4600)$ & $-4758\pm2$ \\
$\sigma_{\rm HARPS}$ & HARPS RV jitter (m s$^{-1}$)& $\mathcal{LU}(10^{-3},100)$ & $9\pm2$ \\
\hline
$q_1^{\rm TESS}$ & TESS linear limb darkening parameter & $\mathcal{U}(0,1)$ & $0.28_{-0.09}^{+0.15}$ \\
$q_2^{\rm TESS}$ & TESS quadratic limb darkening parameter & $\mathcal{U}(0,1)$ & $0.56_{-0.22}^{+0.25}$ \\
$\sigma_{\rm TESS}$ & TESS photometric jitter (ppm) & $\mathcal{LU}(1,5\times10^7)$ & $10_{-8}^{+42}$ \\
$q_1^{r'}$ & MOANA $r'$ linear limb darkening parameter & $\mathcal{U}(0,1)$ & $0.55_{-0.20}^{+0.24}$ \\
$q_2^{r'}$ & MOANA $r'$ quadratic limb darkening parameter & $\mathcal{U}(0,1)$ & $0.62_{-0.32}^{+0.25}$ \\
$\sigma_{r'}$ & MOANA $r'$ photometric jitter (ppm) & $\mathcal{LU}(1,5\times10^7)$ & $17_{-14}^{+114}$ \\
\enddata
\tablecomments{$\mathcal{U}(a,b)$ denotes a uniform prior with a start value $a$ and end value $b$. $\mathcal{G}(\mu,\sigma)$ denotes a normal prior with mean $\mu$, and standard deviation $\sigma$. $\mathcal{LU}(a,b)$ denotes a log-uniform prior with a start value $a$ and end value $b$.}
\end{deluxetable*}

\section{Results and Discussion}\label{sec:discussion}

The Neptune-ridge planet TOI-181\,b was
found to have a misaligned orbit with a sky-projected
spin-orbit angle of $\lambda=32.0_{-6.5}^{+6.3}\,^{\circ}$. As for the orbital eccentricity, \citet{Mistry2023} reported $e=0.15^{+0.06}_{-0.03}$. 
As discussed in Section~\ref{sec:origins},
tidal circularization seems capable of reducing the orbital eccentricity of TOI-181~b to a level well below detectability over the age of the system.
Therefore, we decided to use the $\log{Z}$ as a metric to test for the detectability of eccentricity. A model with $e=0$ is favored over a model
with free parameters for $e$ and $\omega$ by $\Delta\log{Z}=3$. Hence, for subsequent discussion,
we adopted the parameters from the circular model, and a 2$\sigma$ upper limit
$e<0.12$ based on the results of the eccentric model.
Overall, the other planetary parameters resulting
from our analysis are in good agreement with those reported by \citet{Mistry2023}.

For the Neptune-savanna planet TOI-883~b \citep{Rojas2025},
the projected spin-orbit angle was found to
be $\lambda=22_{-14}^{+15}\,^{\circ}$, i.e., we found
no strong evidence for misalignment.
Similar to the previous case, we tested eccentric and circular models. As discussed in Section~\ref{sec:origins}, the tidal circularization timescale is plausibly too long
to reduce the eccentricity below a detectable level
over the age of the system, and indeed the Bayesian
evidence favors the eccentric model by $\Delta\log{Z}=9$. Therefore, for subsequent discussion,
we adopted the parameters from the eccentric model,
including $e=0.16\pm0.03$. The remaining parameters were found to be in good agreement with those reported by \citet{Rojas2025}.

\begin{deluxetable*}{llcr}
\tablecaption{Summary of priors and posteriors of the \texttt{ironman} fit for TOI-883. \label{tab:fit_883}}
\tablewidth{70pt}
\tablehead{Parameter & Description & Prior & Posterior}
\startdata
$\lambda$ & Sky-projected stellar obliquity ($^{\circ}$) & $\mathcal{U}(-180,180)$ & $22_{-14}^{+15}$ \\
$v\sin{i_\star}$ & Projected rotational velocity (km s$^{-1}$) & $\mathcal{U}(0,10)$ & $1.1\pm0.3$ \\
$\rho_\star$ & Stellar density (g cm$^{-3}$) & $\mathcal{G}(1.40,0.15)$ & $1.28_{-0.17}^{+0.14}$ \\
\hline
$P$ & Orbital period (days) & $\mathcal{U}(10.057699,10.057765)$ & $10.057734\pm0.000004$ \\ 
$t_0$ & Transit midpoint (BJD) & $\mathcal{U}(2458466.4681,2458466.4761)$ & $2458466.4721\pm0.0004$ \\
$b$ & Impact parameter & $\mathcal{U}(0,1)$ & $0.66_{-0.05}^{+0.04}$ \\
$i$ & Orbital inclination ($^{\circ}$) & \nodata & $88.2\pm0.2$ \\
$R_p/R_\star$ & Radius ratio & $\mathcal{U}(0,1)$ & $0.066\pm0.001$  \\
$K$ & RV semiamplitude \added{(m s$^{-1}$)} & $\mathcal{U}(0,1000)$ & $12.4\pm0.5$ \\
$\sqrt{e}\sin{\omega}$ & Eccentricity parameter -- sine component & $\mathcal{U}(-1,1)$ & $-0.3\pm0.1$ \\
$\sqrt{e}\cos{\omega}$ & Eccentricity parameter -- cosine component & $\mathcal{U}(-1,1)$ & $0.27\pm0.03$ \\
$e$ & Eccentricity & \nodata & $0.16\pm0.03$ \\
$\omega$ & Argument of periastron ($^{\circ}$) & \nodata & $-47.3_{-7.5}^{+10.8}$ \\
$a/R_\star$ & Scaled semimajor axis & \nodata & $19.0_{-0.9}^{+0.7}$ \\
$R_p$ & Planet radius ($R_\oplus$)& \nodata & $7.13\pm0.13$ \\
$a$ & Semimajor axis (au)& \nodata & $0.087\pm0.004$ \\
$M_p$ & Planet mass ($M_\oplus$)& \nodata & $39.4^{+2.1}_{-2.0}$ \\
$\rho_p$ & Planet density (g cm$^{-3}$)& \nodata & $0.60^{+0.05}_{-0.04}$ \\
\hline
$q_1^{\rm NEID}$ & NEID linear limb darkening parameter & $\mathcal{U}(0,1)$ & $0.57_{-0.35}^{+0.29}$ \\
$q_2^{\rm NEID}$ & NEID quadratic limb darkening parameter & $\mathcal{U}(0,1)$ & $0.56_{-0.37}^{+0.31}$ \\
$\beta$ & Intrinsic stellar line width (km s$^{-1}$) & $\mathcal{G}(4.7,2.0)$ & $4.5_{-1.6}^{+1.9}$ \\
$\gamma_{\rm NEID}$ & NEID RV offset (m s$^{-1}$)& $\mathcal{U}(-100,100)$ & $-0.9\pm0.5$ \\
$\sigma_{\rm NEID}$ & NEID RV jitter (m s$^{-1}$)& $\mathcal{LU}(10^{-3},100)$ & $0.05_{-0.04}^{+0.39}$ \\
$\gamma_{\rm HARPS}$ & HARPS RV offset (m s$^{-1}$)& $\mathcal{U}(16000,17000)$ & $16650\pm1$ \\
$\sigma_{\rm HARPS}$ & HARPS RV jitter (m s$^{-1}$)& $\mathcal{LU}(10^{-3},100)$ & $3\pm1$ \\
$\gamma_{\rm PFS}$ & PFS RV offset (m s$^{-1}$)& $\mathcal{U}(-100,100)$ & $1.6_{-0.4}^{+0.5}$ \\
$\sigma_{\rm PFS}$ & PFS RV jitter (m s$^{-1}$)& $\mathcal{LU}(10^{-3},100)$ & $0.9_{-0.5}^{+0.6}$ \\
\hline
$q_1^{\rm TESS}$ & TESS linear limb darkening parameter & $\mathcal{U}(0,1)$ & $0.31_{-0.10}^{+0.15}$ \\
$q_2^{\rm TESS}$ & TESS quadratic limb darkening parameter & $\mathcal{U}(0,1)$ & $0.16_{-0.12}^{+0.26}$ \\
$\sigma_{\rm TESS}^{\rm 30-min}$ & TESS 30-minute photometric jitter (ppm) & $\mathcal{LU}(1,5\times10^7)$ & $33_{-30}^{+84}$ \\
$\sigma_{\rm TESS}^{\rm 2-min}$ & TESS 2-minute photometric jitter (ppm) & $\mathcal{LU}(1,5\times10^7)$ & $10_{-8}^{+46}$ \\
\enddata
\tablecomments{$\mathcal{U}(a,b)$ denotes a uniform prior with a start value $a$ and end value $b$. $\mathcal{G}(\mu,\sigma)$ denotes a normal prior with mean $\mu$, and standard deviation $\sigma$. $\mathcal{LU}(a,b)$ denotes a log-uniform prior with a start value $a$ and end value $b$.}
\end{deluxetable*}

\subsection{On the Origins of TOI-181\,b and TOI-883\,b}\label{sec:origins}

The proposed mechanisms to explain close-orbiting
exoplanets generally fall into two broad groups: disk-driven evolution and high-eccentricity migration \citep[see, e.g.,][]{Dawson2018}. The disk-driven evolution includes both in-situ formation \citep[e.g.,][]{Batygin2016,Boley2016} and inward migration driven by nebular tides \citep[e.g.,][]{Goldreich1980,Lin1986}. The typical outcome of disk-driven evolution is a planet on an aligned and nearly circular orbit. Recent observations, however, have shown that about a third of protoplanetary disks are misaligned with respect to their host stars \citep{Biddle2025}, calling into question the assumption of primordial star-disk alignment. At the same time,
spin-orbit alignment has been observed in 19 out of the 23 measured systems with a transiting Neptune and at least one
nearby transiting planetary companion (see Section~\ref{sec:isolated_vs_compact} and \citealt{Radzom2024,Radzom2025}), suggestive of primordial alignment.
Although seemingly contradictory, these results might be reconciled if disk-star misalignments primarily affect the outer disk, and preserve good alignment with the inner disk, where transiting planets form and migrate. Indeed, there is already evidence that some disks exhibit strong internal misalignments \citep[e.g.,][]{Marino2015,Ansdell2016,Casassus2018,Francis2020}. In this picture, disk-driven evolution alone is unlikely to account for the misaligned orbit of TOI-181~b. Similarly, the moderately eccentric orbit of TOI-883 b challenges disk-driven evolution scenarios, since disk–planet interactions are expected to damp the eccentricities of close-in sub-Saturn-mass planets on short timescales \citep{Pichierri2024}.
As is the case with many other observed systems,
the observed architectures of these two systems suggest that additional dynamical excitation may have played a role in their evolution.

High-eccentricity tidal migration provides a possible explanation for the observed architectures. In this framework, planets form far away from the star and fall onto highly eccentric orbits as a consequence of planet-planet scattering \citep[e.g.,][]{Rasio1996,Beauge2012}, von Zeipel-Lidov-Kozai oscillations \citep[e.g.,][]{Wu2003,Fabrycky2007,Naoz2011}, or other secular interactions \citep[e.g.,][]{Wu2011,Petrovich2015}, after which tidal dissipation during repeated periastron passages circularizes and shrinks the orbit. In the traditional constant-lag equilibrium-tide theory, the characteristic circularization timescale for pseudo-synchronous rotation  is \citep{Goldreich1966,Ogilvie2007}:
\begin{equation}
\tau_{\rm circ} \equiv -\frac{e}{\dot{e}} = \frac{2P}{63\pi}Q_p^{\prime}\frac{M_p}{M_\star}\left( \frac{a}{R_p} \right)^5 F(e),
\end{equation}
where $Q_p^{\prime}$ is the planet's modified tidal quality factor and $F(e)$ is an eccentricity-dependent correction factor that grows monotonically with increasing  $e$, normalized such that $F(0)=1$ \citep{Hut1981}. Here, $P$, $a$, and $e$ correspond to the initial orbital values. To assess whether this mechanism can reproduce the observed properties of TOI-181~b and TOI-883~b, we integrated the eccentricity evolution assuming that orbital angular momentum is conserved
during tidal circularization and, therefore,
the final orbital distance is $a_f = a(1-e^2)$. For TOI-181~b, we set $a_f$ equal to the observed orbital distance, while for TOI-883~b we set $a_f = a(1-e^2) = 0.086$~au, where $a$ and $e$ are the currently observed values. We consider tidal quality factors in the range $Q_p^{\prime} = 10^4$–$10^5$ \citep[e.g.,][]{Goldreich1966,Correia2020}.

For TOI-181~b, the model indicates that
circularization occurs on timescales of $\sim0.14-1.4$~Gyr, short enough to allow the planet to undergo high-eccentricity tidal migration and be observed today in a nearly circular orbit with $e<0.12$. TOI-883~b presents a more intriguing case, with an estimated
circularization timescale of $\sim2.2-22$~Gyr, of the same
order of magnitude as (or longer than) the estimated
system age of $6.5^{+2.0}_{-2.1}$~Gyr. In the models,
the eccentricity reaches the observed value of 0.16 about $\sim0.7-7$~Gyr after the initial high-eccentricity
excitation. Thus, depending on the value of $Q_p^{\prime}$, the planet could still be in the process of circularization following an earlier phase of high-eccentricity migration.
In short, the derived circularization timescales do not rule out a high-eccentricity migration origin for either system.

Given that the tidal realignment timescales for stars due to Neptune-mass planets are expected to be orders of magnitude longer than the age of the universe \citep[see, e.g.,][]{Zahn1977,Albrecht2012}, it seems reasonable to
suppose that the low stellar obliquities observed for TOI-181~b and TOI-883~b are primordial, rather than the
outcome of tidal evolution. High-eccentricity migration mechanisms are typically associated with misaligned orbits. Nevertheless, a significant number of systems are expected to end up with low obliquities. For example, \citet{Veldhuis2025} showed that $\sim20\%$ of the systems in their von Zeipel-Lidov-Kozai simulations had final obliquities $\lesssim40^{\circ}$ \citep[see also the obliquity distributions presented by][]{Beauge2012,Teyssandier2019}. Furthermore, coplanar high-eccentricity migration \citep{Petrovich2015} could provide a natural explanation for the small obliquities of both systems. Currently, there is no evidence for companions or RV trends in either system. Furthermore, there is no astrometric evidence for
companions, based on the near-unity values of the
Renormalized Unit Weight Error (RUWE) reported in Gaia DR3. Future RV or Gaia DR4 astrometric observations to search for possible massive outer companions and their degree of alignment would be useful to further constrain the formation and evolution of TOI-181~b and TOI-883~b \citep[e.g.,][]{Perryman2014,Espinoza-Retamal2023a,Lammers2025}.

\subsection{POSEIDON Targets in Context}

\citet{Albrecht2021} first identified a possible bimodality in the obliquity distribution. Their sample of true obliquities showed seemingly significant clustering around both $\psi\sim0^\circ$ and $\psi\sim90^{\circ}$. Subsequent studies also identified a large fraction of polar systems \citep[e.g.,][]{Bourrier2023,Attia2023}. However, other analyses using hierarchical Bayesian models for the obliquity distribution
did not find evidence for a pile-up of polar planets \citep{Dong2023,Siegel2023,Espinoza-Retamal2024}. Instead, they concluded that the distribution shows a preference for aligned systems, with a smaller population exhibiting an almost isotropic distribution. While important, these results are based on samples composed almost exclusively of Jupiters on short-period orbits, reflecting the difficulties of measuring obliquities in systems with smaller planets.

Based on a small number of measurements, \citet{Espinoza-Retamal2024} and \citet{Knudstrup2024} found that the preponderance of polar orbits tended to involve Neptune/sub-Saturn planets. If confirmed, this would be a clue about the dynamical history of transiting Neptunes, and how it might differ from that of more massive Jupiters. Our POSEIDON survey aims to understand the dynamical origins of transiting Neptunes. One of the main focuses of the survey is to increase the number of obliquity measurements in order to confirm or rule out the currently observed features in their distribution. 

\begin{figure}
    \centering
    \includegraphics[width=\linewidth]{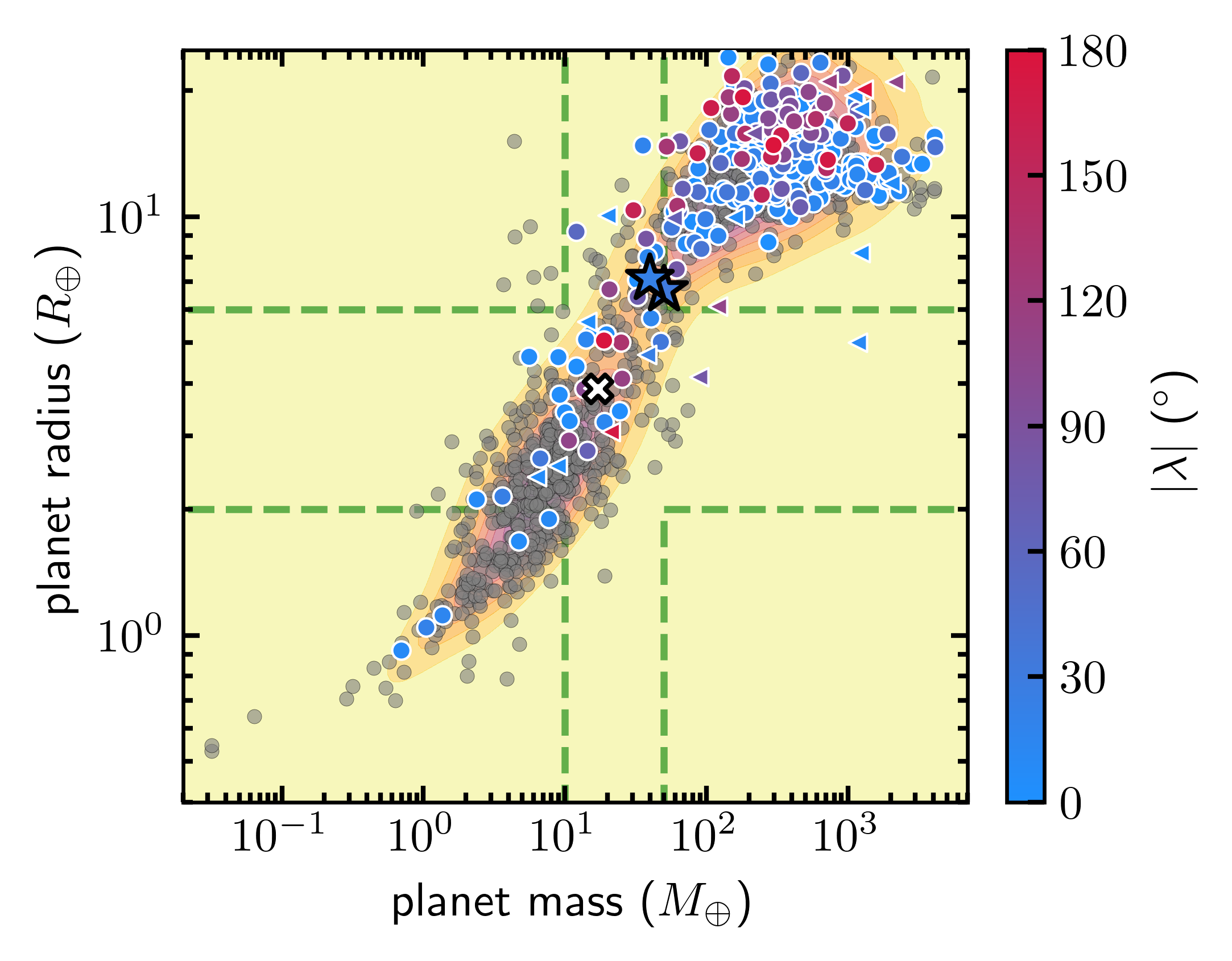}
    \caption{Mass versus radius diagram. Grey points are all the transiting exoplanets from TEPCat \citep{Southworth2011} as of November 2025. Colored points convey the projected obliquity for systems with published measurements. Triangles indicate systems with only upper limits available for the mass. TOI-181~b and TOI-883~b are highlighted as stars following the same color code. The green dashed lines indicate the region where $10\leq M_p/M_\oplus \leq 50$ or $2\leq R_p/R_\oplus \leq 6$, which we considered as Neptunes. Neptune itself is highlighted as the white cross. Contours show the density of points.}
    \label{fig:mass_vs_radius}
\end{figure}

\begin{figure*}
    \centering
    \includegraphics[width = \textwidth]{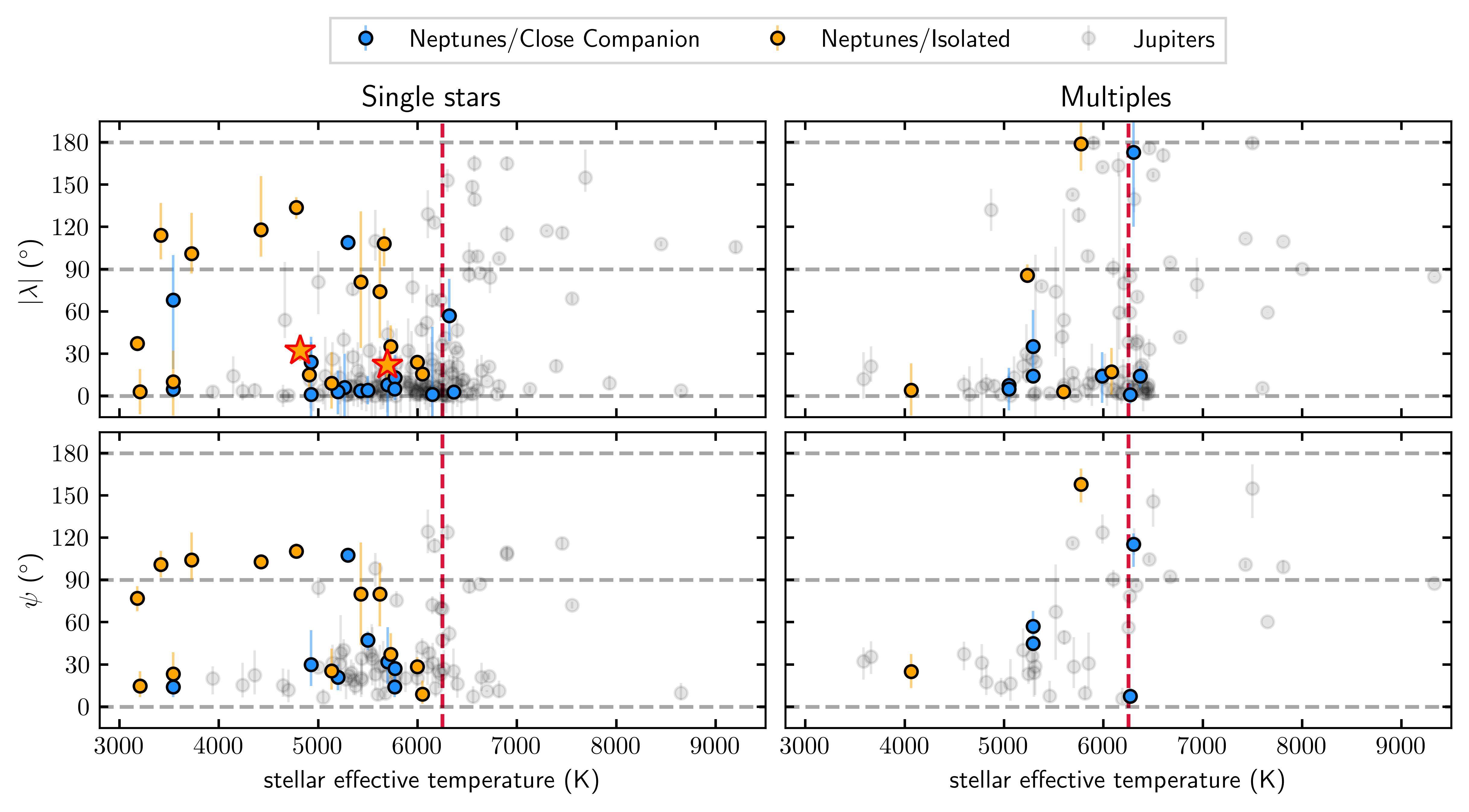}
    \caption{Stellar obliquity of Neptune ($2\leq R_p/R_\oplus\leq 6$ or $10\leq M_p/M_\oplus\leq 50$) and Jupiter (not Neptune with $R_p/R_\oplus > 6$) hosts as a function of the stellar effective temperature. Upper panels show the projected obliquity while lower panels the true obliquity. Left panels show systems where no companion star has been reported and right panels show known multiple-star systems. Neptunes that have planetary companions with periods between 1/5th and 5 times the period of the Neptune are shown in blue, while those without such close companions are shown in orange. Measurements performed in this work for TOI-181 and TOI-883 are highlighted as stars with a red edge, and follow the same color code. Jupiters are shown in the background in black. Literature measurements come from \citet{Rossi2025} and TEPCat \citep{Southworth2011}. The dashed red line demarks the Kraft break at $T_{\rm eff}\approx6250$~K \citep{Kraft1967}.}
    \label{fig:obl_vs_teff}
\end{figure*}

The fact that neither TOI-181~b nor TOI-883~b was found to have a polar orbit weakens the evidence for a polar peak in the distribution. Following \citet{Espinoza-Retamal2024}, we define Neptunes as planets with $10\leq M_p/M_\oplus \leq 50$ or $2\leq R_p/R_\oplus \leq 6$ (see Figure~\ref{fig:mass_vs_radius}). While our primary cut is based on mass, which is the primary driver of the dynamics, we also included a cut on size to consider planets with poorly constrained masses or only upper limits available, but that given their small sizes, are likely part of the same population. To create a sample of Neptune systems with measured obliquities from the literature for comparison and statistical analyses, we started from the recent catalog published by \citet{Rossi2025}. This catalog reports true obliquities homogeneously derived from the reported projected values with the method of \citet{Masuda2020}. Several systems are missing from the catalog, as it only includes systems with a reported stellar rotation period and does not consider cases where this missing information precludes the true obliquity measurement. To account for these missing systems, we merged the catalog with the stellar obliquity information available in TEPCat\footnote{\url{https://www.astro.keele.ac.uk/jkt/tepcat/}} \citep{Southworth2011} as of November 2025. Additionally, we  included in our sample the measurement for TOI-1710\,b ($\lambda=179\pm19^{\circ}$, $\psi=158_{-13}^{+11}\,^{\circ}$), \added{the subject of the second POSEIDON paper \citep{Espinoza-Retamal2026}}, and a new and more precise measurement for WASP-156\,b ($\lambda=-15^{+17}_{-16}\,^{\circ}$) that will be described in separate publication \added{\citep{Lafarga2026}}. By doing so, we ended up with a sample of 43 Neptunes from the literature, plus the two reported here for a total of 45 Neptunes \added{(see Appendix~\ref{app:sample} for a complete list of the Neptunes and their properties)}. Figure \ref{fig:obl_vs_teff} shows the obliquity values for Neptune hosts as a function of the stellar effective temperature. \added{It is important to note that almost all Neptune systems with obliquity measurements have stars with $T_{\rm eff}\lesssim6250$~K, i.e.,
below the Kraft break \citep{Kraft1967}. For close-in giant planets, the obliquity distributions of stars above and below the Kraft break
have been found to differ 
\citep{Schlaufman2010,Winn2010,Wang2026}, with hot stars exhibiting a broader range of obliquities than cool stars. The lack of obliquity measurements for Neptune hosts above the Kraft break may reflect the observational difficulty of detecting and confirming small transiting planets around hot stars.} 

\subsection{The True Obliquity Distribution}\label{sec:HBM}

\begin{deluxetable*}{lc|ccc|ccc|r}
\tablecaption{Summary of posteriors for the parameters of the derived $\cos{\psi}$ distribution. We modeled the distribution as the combination of two Beta distributions, each with relative weight $w$, mean $\mu$, and variance $1/\kappa$. For larger values of $\kappa$ the distribution is more concentrated around the mean. Since the Beta distribution is defined between 0 and 1, the peak of each component is located at $\cos{\psi}=2\mu-1$.\label{tab:HBM}}
\tablecolumns{9}
\tablewidth{0pt}
\tablehead{ & & \multicolumn{3}{c|}{Misaligned Component} & \multicolumn{3}{c|}{Aligned Component} & \\ 
Sample & Systems & $w_0$ & $\mu_0$ & $\log\kappa_0$ & $w_1$ & $\mu_1$ & $\log\kappa_1$ & Reference
}
\startdata
{\bf Jupiters} &  &  &  &  &  &  &  & \\
All & 235 & $0.43_{-0.10}^{+0.12}$ & $0.63_{-0.10}^{+0.08}$ & $0.01_{-0.23}^{+0.31}$ & $0.57_{-0.12}^{+0.10}$ & $0.99\pm0.01$ & $4.88_{-0.87}^{+1.18}$ & This work\\
Cool Star & 150 & $0.26^{+0.10}_{-0.08}$ & $0.65^{+0.10}_{-0.14}$ & $0.15^{+0.42}_{-0.36}$ & $0.74_{-0.10}^{+0.08}$ & $0.99\pm0.01$ & $4.94^{+2.76}_{-1.11}$ & This work\\
Hot Star & 85 & $0.12_{-0.08}^{+0.43}$ & $0.43_{-0.17}^{+0.16}$ & $1.60_{-1.88}^{+3.93}$ & $0.88_{-0.43}^{+0.08}$ & $0.77_{-0.05}^{+0.17}$ & $-0.17_{-0.26}^{+1.67}$ & This work\\
\hline
{\bf Neptunes} &  &  &  &  &  &  &  & \\
All & 27 & $0.39_{-0.31}^{+0.24}$ & $0.40_{-0.14}^{+0.12}$ & $4.61_{-3.63}^{+5.15}$ & $0.61_{-0.24}^{+0.31}$ & $0.94_{-0.23}^{+0.05}$ & $1.61_{-2.08}^{+5.61}$ & JIER2024\tablenotemark{a}\\
All & 45 & $0.22^{+0.14}_{-0.12}$ & $0.36^{+0.19}_{-0.10}$ & $2.32^{+3.23}_{-2.04}$ & $0.78_{-0.14}^{+0.12}$ & $0.96^{+0.02}_{-0.09}$ & $2.32^{+1.27}_{-1.87}$ & This work\\
\hline
Isolated & 22 & $0.31^{+0.28}_{-0.20}$ & $0.35^{+0.22}_{-0.16}$ & $1.83^{+4.37}_{-1.84}$ & $0.69^{+0.20}_{-0.28}$ & $0.88^{+0.06}_{-0.13}$ & $1.54^{+3.41}_{-1.28}$ & This work\\
Close companion & 23 & $0.12^{+0.09}_{-0.06}$ & $0.36^{+0.10}_{-0.08}$ & $4.49^{+3.73}_{-3.77}$ & $0.88^{+0.06}_{-0.09}$ & $0.99^{+0.01}_{-0.01}$ & $5.40^{+2.57}_{-2.19}$ & This work\\
\hline
Single stars & 32 & $0.21^{+0.10}_{-0.09}$ & $0.34\pm0.06$ & $5.79^{+2.92}_{-2.99}$ & $0.79^{+0.09}_{-0.10}$ & $0.95^{+0.02}_{-0.05}$ & $2.29^{+0.99}_{-1.37}$ & This work\\
Multiples & 13 & $0.27^{+0.21}_{-0.15}$ & $0.32^{+0.23}_{-0.20}$ & $-0.11^{+2.81}_{-1.32}$ & $0.73^{+0.15}_{-0.21}$ & $0.98^{+0.01}_{-0.12}$ & $3.72^{+4.08}_{-3.61}$ & This work\\
\hline
Hot ($P<5.7$~d) & 15 & $0.22^{+0.20}_{-0.15}$ & $0.33^{+0.21}_{-0.14}$ & $3.71^{+4.24}_{-3.22}$ & $0.78^{+0.15}_{-0.20}$ & $0.86^{+0.08}_{-0.09}$ & $0.36^{+0.97}_{-0.70}$ & This work\\
Warm ($P>5.7$~d) & 30 & $0.20^{+0.18}_{-0.13}$ & $0.38^{+0.23}_{-0.19}$ & $0.97^{+3.66}_{-1.63}$ & $0.80^{+0.13}_{-0.18}$ & $0.97^{+0.02}_{-0.11}$ & $2.41^{+2.19}_{-2.23}$ & This work\\
\enddata
\tablecomments{$^a$ \citet{Espinoza-Retamal2024}}
\end{deluxetable*}

In order to derive the underlying stellar obliquity distribution of Neptune hosts, including the two values reported here, and compare it to the sample of Jupiter hosts, we applied the Bayesian framework presented by \citet{Dong2023}. This framework models the distribution of $\cos{\psi}$ across an exoplanet population using a mixture model of Beta distributions \citep[e.g.,][]{Gelman14}, which can capture a broad range of distributions from isotropic to strongly multi-modal, depending on the number of Beta components assumed. Although the framework can work with different observables to derive the $\cos{\psi}$ distribution, here we derived it only from the $\lambda$ measurements and did not include information about the stellar inclination.
Each Beta distribution is described by 3 parameters: $w$, $\mu$, and $\kappa$. The parameter $w$ describes the weight of the component (the fraction of systems belonging to that
component), while $\mu$ and $1/\kappa$ are the mean and variance. The $\mu$ and $\kappa$ parameters can be related to the typical $\alpha$ and $\beta$ parameters of a Beta distribution as $\alpha = \mu \kappa$ and $\beta=(1-\mu)\kappa$. In all cases, we assumed uniform priors between 0 and 1 for $\mu$ and between $-4$ and 10 for $\log\kappa$. Table \ref{tab:HBM} shows the results obtained from the posterior distributions for the different populations discussed below.

\subsubsection{All Jupiters}

We first applied the Bayesian framework to derive the obliquity distribution for the sample of transiting Jupiters. We considered Jupiters to be planets that do not qualify as Neptunes and have $R_p/R_\oplus>6$ (the population in the upper right region of Figure \ref{fig:mass_vs_radius}). This will allow comparison with the distribution of Neptune hosts. As shown in Figure \ref{fig:old_vs_new}, the obliquity distribution for Jupiter hosts shows a preference for aligned orbits, and there is a second population with nearly isotropic orientations and no significant clustering near $90^\circ$.  This is in good agreement with results reported previously \citep[e.g.,][]{Dong2023,Siegel2023,Espinoza-Retamal2024,Espinoza-Retamal2025,Knudstrup2024}, but is more precisely constrained thanks to the updated sample of 235 Jupiters. The relative weights of the two Beta components are $w_0\approx0.4$ and $w_1\approx0.6$, suggesting that $\sim 60\%$ of Jupiter hosts have low obliquities and the remaining $\sim 40\%$ follow a nearly isotropic stellar obliquity distribution. The explanation for the isotropic population
could be high-eccentricity migration, which typically results in broad ranges of obliquities and can also account for the
absence of nearby planetary companions
to hot Jupiters \cite[their ``loneliness''; see, e.g.,][]{Becker2025,Sha2026}. On the other hand, there are many reasons a system might be observed to be aligned: formation in an aligned disk (see the discussion in Section \ref{sec:origins}), coplanar high-eccentricity migration \citep{Petrovich2015}, or tidal obliquity damping which is non-negligible for Jupiter hosts \citep[e.g.,][]{Winn2010,Albrecht2022}.

\subsubsection{Jupiters: Cool versus Hot host star}

\added{The obliquity
distribution of stars with transiting giant planets has been shown to vary with effective temperature and, in particular, on whether the host star lies above or below the Kraft break. In general, hot stars exhibit a broader range of obliquities than cooler stars \citep{Schlaufman2010,Winn2010,Albrecht2012,Wang2026}. Motivated by these results, we examined the stellar obliquity distributions separately for Jupiter hosts above and below the Kraft break.

As shown in Figure~\ref{fig:old_vs_new}, both distributions exhibit the same overall structure as the combined Jupiter population. In both cases, most systems are aligned, and there is a second population that exhibits nearly isotropic obliquities. The main difference between cool and hot stars lies in the degree of concentration of the aligned component. For cool stars, the aligned population is sharply peaked, with $\log\kappa_1 = 4.94^{+2.76}_{-1.11}$, while for hot stars the aligned component is broader, with $\log\kappa_1 = -0.17^{+1.67}_{-0.26}$. This difference
suggests that, when considering the
aligned components of each distribution,
the cool stars are more strongly aligned
than the hot stars.

This result is consistent with the expectation that tidal dissipation in the convective envelopes of cool stars can damp stellar obliquities over time, producing a more sharply aligned population among Jupiter hosts below the Kraft break \citep[e.g.,][]{Winn2010,Albrecht2012,Rice2022a,Zanazzi2024b}. In contrast, tidal realignment is expected to be less efficient for hotter stars with thinner convective envelopes, allowing a broader range of obliquities to persist.}

\subsubsection{All Neptunes}

\begin{figure*}
    \centering
    \includegraphics[width=\linewidth]{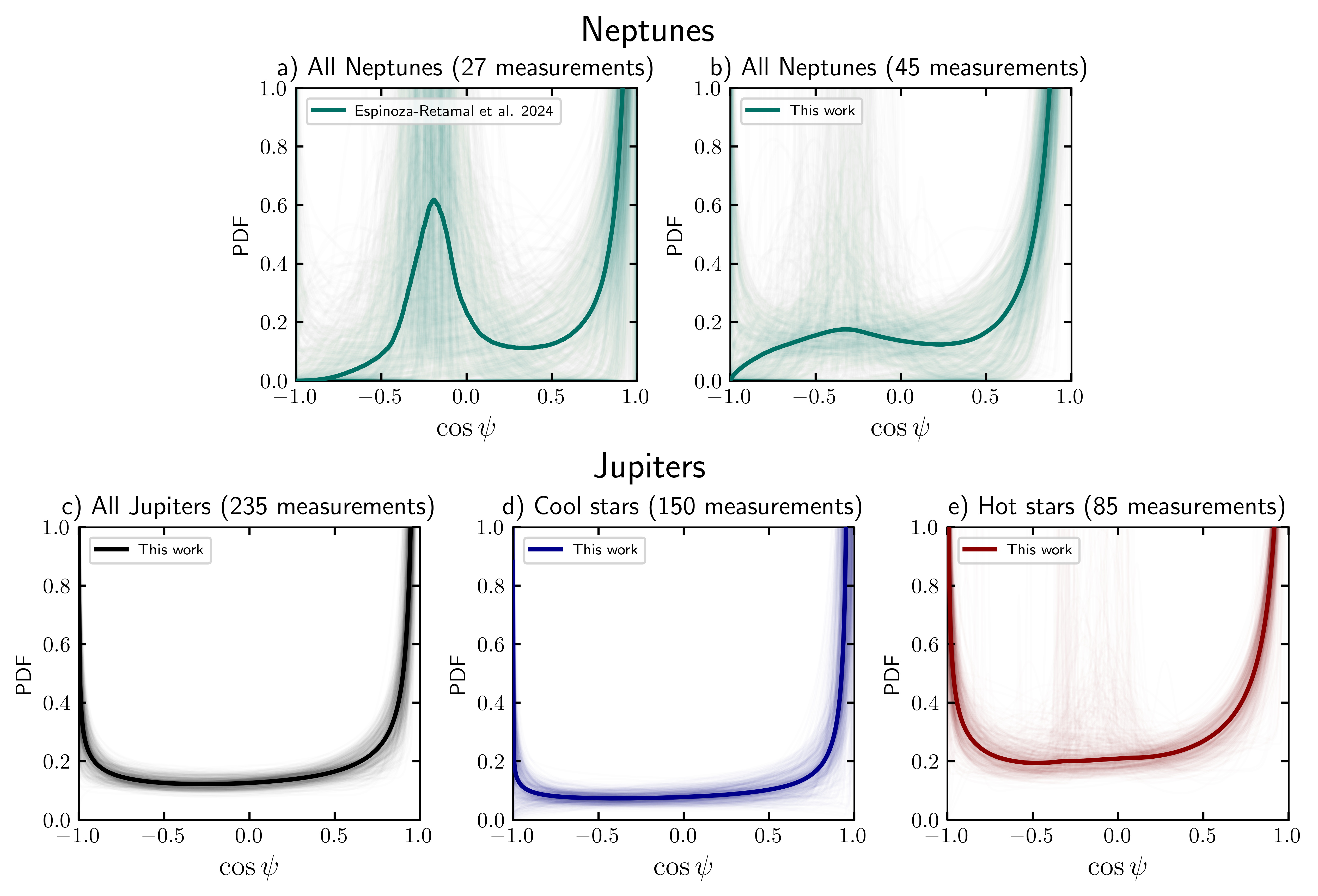}
    \caption{Inferred stellar obliquity distributions for the samples of Jupiter and Neptune hosts. This inference was based on sky-projected obliquity measurements, following the methodology of \citet{Dong2023}, without including information about the stellar inclination. The solid line shows the median distribution, and random samples from the posteriors are shown in the background. a) Distribution derived by \citet{Espinoza-Retamal2024} for a sample of 27 Neptune host stars  (green). b) Distribution derived in this work for the expanded sample of 45 Neptune host stars (green). c) Distribution derived in this work for the sample of 235 Jupiter host stars (black). d) Distribution derived in this work for the sample of 150 Jupiter host stars with $T_{\rm eff}\leq6250$~K (blue). e) Distribution derived in this work for the sample of 85 Jupiter host stars with $T_{\rm eff}>6250$~K (blue).}
    \label{fig:old_vs_new}
\end{figure*}

We then applied the Bayesian framework to derive the obliquity distribution for the entire sample of 45 Neptune hosts. This allows a direct comparison with the distributions derived in \citet{Espinoza-Retamal2024}, who performed the same analysis but for a smaller sample of 27 Neptunes, and with the population of Jupiter hosts. As shown in Figure \ref{fig:old_vs_new}, the distribution derived here appears different than the one derived with a smaller sample. The key difference between them is the significance of the polar peak. \citet{Espinoza-Retamal2024} found $w_0\approx0.4$ and $w_1\approx0.6$, suggesting that $\sim60\%$ of the Neptunes were aligned and $\sim40\%$ were polar. With the larger sample, here we found $w_0\approx0.2$ and $w_1\approx0.8$, suggesting that only $\sim20\%$ of the Neptunes have nearly polar orbits. Further, \citet{Espinoza-Retamal2024} found $\log{\kappa_0}=4.61^{+5.15}_{-3.63}$ while we found $\log{\kappa_0}=2.32^{+3.23}_{-2.04}$, meaning that the new distribution is less concentrated around 90$^\circ$.

These differences are a consequence of the fact that most of the new Neptune systems added to the sample are aligned, and according to our definition of Neptunes, no new polar Neptunes have been reported. Furthermore, although WASP-156~b was originally reported as a polar planet by \citet{Bourrier2023}, more precise observations by \citet{Lafarga2026} indicate that the orbit is actually well-aligned.

\added{The updated obliquity distribution of Neptune host stars resembles that observed for Jupiter host stars, with an aligned population together with a second population exhibiting nearly isotropic obliquities (see Figure~\ref{fig:old_vs_new}). However, the aligned component for Neptune hosts appears broader than that observed for Jupiter hosts below the Kraft break. Interestingly, the Neptune distribution is more similar to that observed for Jupiter hosts above the Kraft break. One possible interpretation is that transiting Neptunes and Jupiters share similar primordial obliquity distributions, but tidal realignment is more efficient for Jupiter hosts due to their larger planetary masses \citep[e.g.,][]{Zahn1977,Albrecht2012}. In this picture, the narrower aligned distribution observed for Jupiter hosts below the Kraft break would result from subsequent tidal damping of stellar obliquities, while Neptune hosts retain a broader aligned component because tides are less effective.
A more direct comparison would
be between hot stars with Neptunes and
hot stars with Jupiters, but 
we do not yet have a sample of obliquity
measurements of hot stars with Neptunes
(see Figure~\ref{fig:obl_vs_teff}). 
}

\subsubsection{Neptunes: Isolated versus Close companion}\label{sec:isolated_vs_compact}

We also derived the obliquity distribution for the samples of Neptunes that have close companions (i.e., those in compact multi-planet systems) and those for which no close
companions have been detected (apparently ``isolated'' planets). One might expect
these samples to differ because the architectures of compact multi-planet systems are thought to be shaped by the disk \citep[see, e.g.,][]{Radzom2024,Radzom2025,Polanski2025}, whereas more isolated planets can also be consistent with high-eccentricity migration. \citet{Polanski2025} already
noted that these samples have different distributions
of planetary radius, with isolated Neptunes being larger and typically more misaligned than those with close companions. Here, we considered as close companions planets that have periods between 1/5th and 5 times the period of the Neptune. As most of the Neptunes in our sample have been discovered by TESS, which offers limited sensitivity to longer-period planets, we acknowledge that some of the planets we label as isolated could actually be part of compact multi-planet systems. However, the current data are insufficient to confirm the presence of close companions.

\begin{figure*}
    \centering
    \includegraphics[width=0.75\linewidth]{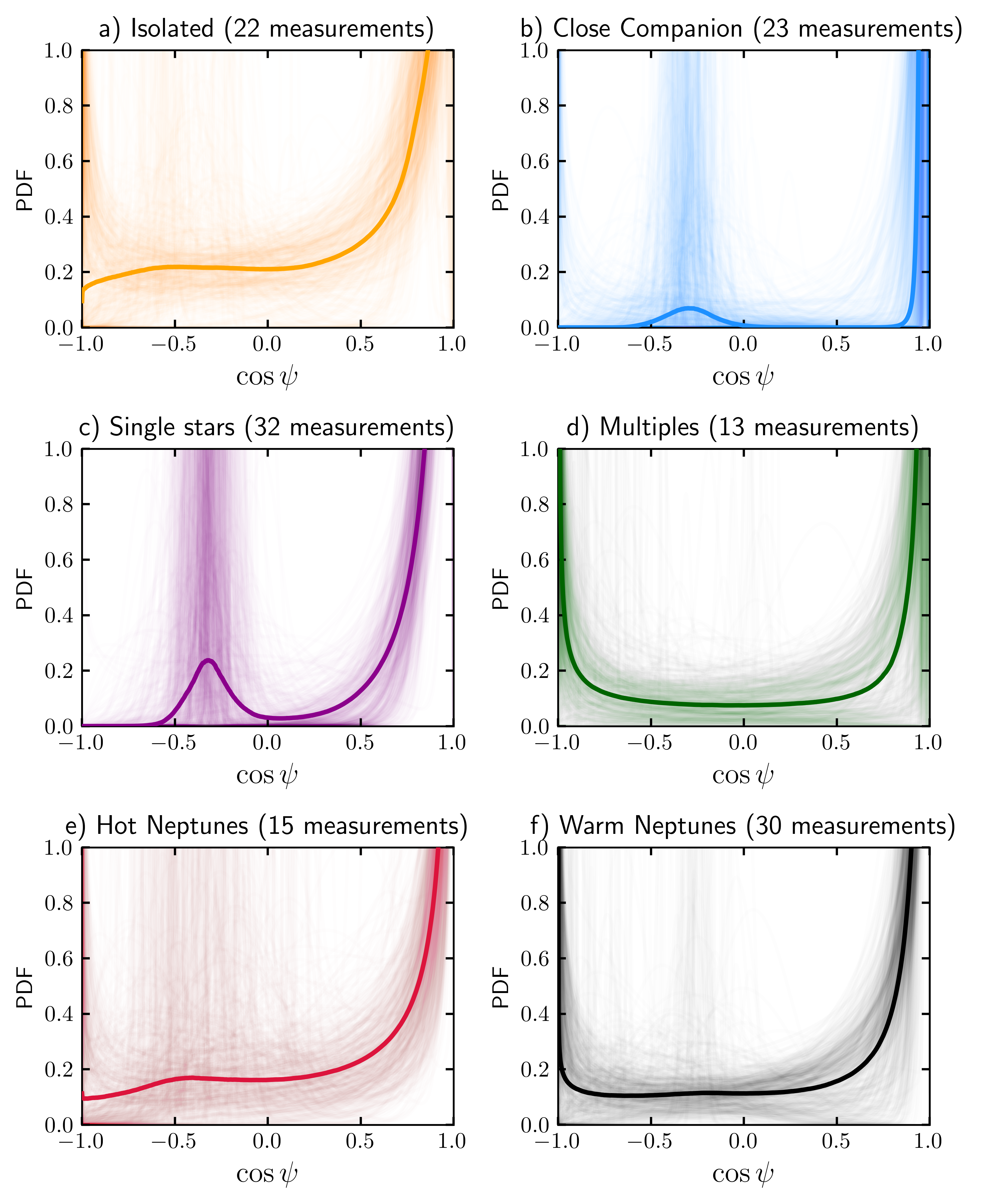}
    \caption{Inferred stellar obliquity distributions for the different subsamples of Neptune hosts. In all panels the median distribution is shown as the solid line, and random samples from the posteriors are shown in the background. a) Neptunes that are apparently isolated, i.e., that
    have no known close planetary companions (orange). b) Neptunes with known close companions (blue). c) Neptune hosts that are single stars (purple). d) Neptune hosts that are part of multi-star systems (green). e) Hot Neptunes (red). f) Warm Neptunes (black).}
    \label{fig:distributions}
\end{figure*}

As shown in Figure \ref{fig:distributions}, we found the $\cos{\psi}$ distribution for compact multi-planet systems to be consistent with a population of almost perfectly aligned systems, and some excess at $\cos{\psi}\approx-0.28$ or $\psi\approx106^{\circ}$. Neptunes with known close companions are usually very well-aligned, with some notable exceptions (see Figure~\ref{fig:obl_vs_teff}). This suggests that the parts of protoplanetary disks where these planets form and evolve are preferentially aligned with the stellar equator. The four exceptions for which high obliquities have
been reported ($|\lambda|>35^{\circ}$) are HIP~41378 \citep{Grouffal2022,Grouffal2025}, AU~Mic \citep{Hirano2020,Palle2020,Addison2021,Martioli2021,Yu2025}, HD~3167 \citep{Dalal2019,Bourrier2021}, and K2-290 \citep{Hjorth2021}. For three of these systems,
we consider the observational evidence to
be less than definitive. For HIP~41378, this is because
the RM observations were performed for long-period planets, for which the long
transit durations made it necessary
to observe different parts of the transit
with different instruments. For AU~Mic, the RM data
were strongly affected by stellar activity and have
large uncertainties; furthermore, it is not clear
that the reported system parameters are compatible
with long-term dynamical stability. For HD~3167, the RM signal of the inner planet was detected with a low-signal-to-noise ratio. More observations of these systems are required for conclusive results. For the remaining system, K2-290, the finding of a large misalignment seems secure. \citet{Hjorth2021} proposed that the misalignment
was primordial; more recently \citet{Best2022}
provided an alternative interpretation in which
the system was primordially aligned and 
become misaligned due to the influence of the two known stellar companions.

In contrast, the population of apparently isolated Neptunes seems to have a different obliquity distribution. The mixture model consists of a large fraction of aligned systems and
a smaller population of systems with nearly
random orientations, with no detectable clustering at $90^\circ$ (see Figure~\ref{fig:distributions}). Similar to the entire population of Neptunes, the distribution resembles that of transiting Jupiters, suggesting that Jupiters and isolated Neptunes have a common origin. This is further supported by the fact that both populations are preferentially hosted by metal-rich stars \citep{Dong2018}.
For the population with apparently randomly oriented host stars, high-eccentricity tidal migration is
an appealing explanation that can also account for their loneliness \cite[e.g.,][]{Becker2025,Sha2026}. Similarly, as discussed in Section \ref{sec:origins}, the aligned and isolated Neptunes can, in principle, be explained by coplanar high-eccentricity migration \citep{Petrovich2015}, although formation and evolution within aligned disks is also possible. Future follow-up observations to constrain outer companions \citep[e.g., using Gaia astrometry;][]{Perryman2014,Espinoza-Retamal2023a,Lammers2025} will be helpful to estimate the fraction of aligned Neptunes produced by disk migration versus coplanar high-eccentricity migration.

\subsubsection{Neptunes: Single stars versus Multiples}

We also derived the obliquity distribution for the samples of Neptunes in single-star and multi-star systems. These two samples might be expected to differ because the presence of additional stars in the system can alter the initial conditions and make certain dynamical scenarios more or less likely to occur. To distinguish between single stars and multiples, we made use of the information available in the Encyclopedia of Exoplanetary Systems\footnote{\url{https://exoplanet.eu/}} and its catalog of binary systems presented by \citet{Thebault2025}. We ended up with 32 Neptunes around single stars and 13 in binary systems. We note that 12 of the 13 binaries were identified by \citet{El-Badry2021} using the proper motions and parallaxes from the Gaia DR3, implying that they are very wide binaries with separations of 200--10000 au. The only system not identified in this way is HD~148193, for which \citet{Chontos2024} reported a binary M dwarf companion at a physical separation of $\sim140$ au found using high-resolution imaging. As usual in these types of
comparisons, we must keep in mind that
some of the stars classified here as ``single'' might have stellar companions that have not been identified with the current data. 

As shown in Figure \ref{fig:distributions}, we found that for Neptunes that orbit single stars, the polar peak is more significant than in all the other cases we considered, although with large uncertainties due to the small number of measurements (see Table~\ref{tab:HBM}). Neptunes cluster around $\cos{\psi}\approx-0.3$ or $\psi\approx108^{\circ}$ with a concentration $\log{\kappa_0}=5.79^{+2.92}_{-2.99}$. This result was expected, as all of the known polar Neptunes orbit single stars except WASP-139~b \citep[][; see Figure~\ref{fig:obl_vs_teff}]{Espinoza-Retamal2024}. A possible explanation for this preference for polar orbits is the disk-driven resonance mechanism proposed by \citet{Petrovich2020} and further extended by \citet{Louden2024} and \citet{Zanazzi2024}. This mechanism relies on the interaction between the close-in planet, an outer planetary companion, and an evaporating protoplanetary disk. A resonance between the precession frequencies of the planet's orbit induced by the disk and by the outer companion excites the planet's orbital inclination, and if the general relativistic precession rate is fast enough, the inclination is driven toward $90^\circ$. \citet{Handley2026} proposed a similar mechanism that relies on opening a gap in the early protoplanetary disk rather than invoking
an outer companion. Another possible
explanation for this polar peak is high-eccentricity tidal migration induced by von Zeipel-Lidov-Kozai oscillations. This mechanism is expected to produce a peak at $\psi\approx115^{\circ}$ \citep{Fabrycky2007}, although the significance and exact position of the peak depend on several unknown parameters \citep[e.g.,][]{Naoz2011,Anderson2016}. More detailed simulations would be interesting to see if this scenario could be responsible for the polar orbits of Neptunes around single stars.

In contrast, the obliquity distribution for Neptune hosts in known multiple-star systems has a preference for aligned systems with a smaller population that has almost isotropic orientations and a small excess of retrograde orbits (see Figure~\ref{fig:distributions}). The fact that only three out of the 13 systems are misaligned ($|\lambda|>35^{\circ}$; see Figure~\ref{fig:obl_vs_teff}) suggests that the known companions are not playing important roles in producing misaligned Neptunes. The apparent excess of retrograde
systems is based on only three systems: K2-290 \citep{Hjorth2021}, TOI-1710 \citep{Espinoza-Retamal2026}, and WASP-139 \citep{Espinoza-Retamal2024}.
As such, we refrain from attempting an interpretation
until the sample can be enlarged.

\subsubsection{Neptunes: Hot Neptune versus Warm Neptune}

As a final test, we derived the underlying obliquity distribution for the samples of hot and warm Neptune hosts to see if there are differences in the obliquity distribution of Neptunes that are in the desert, ridge, and savanna (see Figure \ref{fig:neptune_desert}). We made the cut between hot and warm at an orbital period of 5.7 d, which was
found by \cite{Castro-Gonzalez2024} to separate the ridge from the savanna.

As shown in Figure \ref{fig:distributions}, there is no detectable difference between the obliquity distributions of hot and warm Neptune hosts. Both distributions are consistent with an aligned population and a smaller population with an almost isotropic distribution of spin-orbit angles. The distribution of hot Neptune hosts shows little evidence for a clustering at around $\cos{\psi}\approx-0.34$ or $\psi\approx110^{\circ}$ (see Table~\ref{tab:HBM}). However, the sample consists of only 15 systems. More measurements would allow for a more robust comparison.

\section{Summary and Conclusions}\label{sec:conclusion}

In this first paper of the POSEIDON survey, we presented RM observations and stellar obliquity measurements for two systems, TOI-181 and TOI-883. In addition to characterizing these individual systems, we performed a statistical analysis of the stellar obliquity distribution of transiting Neptunes using an updated sample from the literature. We concluded that:

\begin{itemize}
    \item TOI-181~b is a Neptune on a low-eccentricity orbit ($e<0.12$ at $2\sigma$) with a modest but statistically significant spin-orbit misalignment $\lambda=32.0^{+6.3}_{-6.5}\,^{\circ}$. The misalignment is suggestive of high-eccentricity migration.

    \item TOI-883~b shows no securely detectable spin-orbit
    misalignment ($\lambda = 22^{+15}_{-14}\,^{\circ}$)
    but does have a detectable eccentricity $e = 0.16 \pm 0.03$. High-eccentricity migration followed by incomplete tidal circularization offers a viable explanation for this system. Coplanar high-eccentricity migration, in particular, can preserve low obliquities.

    \item The evidence for a bimodal obliquity distribution for Neptune hosts has been weakened after the addition of new systems to the sample. Instead, the observed distribution shows a dominant aligned population and a smaller population with nearly isotropic orientations. This updated distribution resembles that of transiting Jupiters, suggesting that both populations share common formation and migration pathways. \added{However, the aligned component for Neptune hosts appears broader than that observed for Jupiter hosts below the Kraft break, consistent with the possibility that tidal realignment is less efficient for Neptune hosts.}

    \item Neptunes in compact multi-planet systems are usually aligned, consistent with formation and migration within aligned disks. In contrast, transiting Neptunes without any detected close companions exhibit a broader, nearly isotropic obliquity distribution, suggestive of a larger role for dynamical excitation and high-eccentricity migration.

    \item Neptunes orbiting single stars still show
    tentative evidence for a peak in the obliquity
    distribution near 90$^\circ$, with large error bars due to the small number of measurements. Neptunes in known multi-star systems do not display evidence for a polar peak. Although current uncertainties remain large, this result suggests that the presence of stellar companions may alter the efficiency or outcomes of the mechanisms that produce polar orbits.
    
\end{itemize}

The POSEIDON survey is designed to substantially expand the sample of Neptune hosts with obliquity measurements, enabling more precise population-level constraints and a clearer assessment of how planet mass, multiplicity, and stellar environment influence the dynamical histories of transiting exoplanets.

\begin{acknowledgments}

We thank Alessandro Rossi for kindly sharing his recently published catalog of stellar obliquities. We also thank Xian-Yu Wang for his help with the scheduling of the NEID observations. \added{We would also like to thank the anonymous referee for their thoughtful review and suggestions that improved the quality of this work.}

This work is based on observations collected with the Magellan Clay Telescope at Las Campanas Observatory, Chile, under the programme allocated by the Chilean Telescope Allocation Committee (CNTAC) number CN2025A-9 (PI Juan I.\ Espinoza-Retamal).

This work is based on observations taken with the NEID instrument on the WIYN 3.5 m telescope at Kitt Peak National Observatory (Proposal ID 2025B-490658, PI Joshua Winn). We thank the NEID Queue Observers and WIYN Observing Associates for their skillful execution of our NEID observations. The authors are honored to be permitted to conduct astronomical research on I'oligam Du'ag (Kitt Peak), a mountain with particular significance to the Tohono O'odham. Kitt Peak is a facility of NSF's NOIRLab, managed by the Association of Universities for Research in Astronomy (AURA). The WIYN telescope is a joint facility of NOIRLab, Indiana University, the University of Wisconsin-Madison, Pennsylvania State University, Purdue University, and Princeton University. NEID was funded by the NASA-NSF Exoplanet Observational Research (NN-EXPLORE) partnership and built by Pennsylvania State University. The NEID archive is operated by the NASA Exoplanet Science Institute at the California Institute of Technology. NN-EXPLORE is managed by the Jet Propulsion Laboratory, California Institute of Technology under contract with the National Aeronautics and Space Administration. 

This paper includes data collected with the TESS mission, obtained from the MAST data archive at the Space Telescope Science Institute (STScI). Funding for the TESS mission is provided by the NASA Explorer Program. STScI is operated by the Association of Universities for Research in Astronomy, Inc., under NASA contract NAS 5–26555.

This research has made use of data obtained from or tools provided by the portal \href{https://exoplanet.eu/}{exoplanet.eu} of The Extrasolar
Planets Encyclopedia.

\added{J.N.W.\ acknowledges a grant associated with NEID observing programs from the NN-EXPLORE program (JPL RSA 1719095).}

R.B.\ acknowledges support from Fondecyt Project 1241963

A.J.\ acknowledges support from Fondecyt project 1251439.

This work has been carried out within the framework of the National Centre of Competence in Research PlanetS supported by the Swiss National Science Foundation (SNSF) under grant 51NF40\_205606. M.H.\ acknowledges the financial support of the SNSF.

\facilities{Magellan:Clay (PFS), WIYN (NEID), El Sauce: 0.6m, TESS, ESO:3.6m (HARPS), MAST}

\software{
\texttt{astropy} \citep{astropy,astropy2,astropy3},
\texttt{batman} \citep{batman},
\texttt{celerite} \citep{celerite},
\texttt{dynesty} \citep{dynesty2},
\texttt{ironman} \citep{Espinoza-Retamal2023b,Espinoza-Retamal2024},
\texttt{juliet} \citep{juliet},
\texttt{lightkurve} \citep{lightkurve},
\texttt{matplotlib} \citep{matplotlib},
\texttt{numpy} \citep{numpy},
\texttt{radvel} \citep{Fulton18},
\texttt{rmfit} \citep{Stefansson2022},
\texttt{scipy} \citep{scipy},
\texttt{serval} \citep{Zechmeister2018},
\texttt{zaspe} \citep{zaspe}.
}

\end{acknowledgments}

\bibliography{sample7}{}
\bibliographystyle{aasjournalv7}

\appendix
\restartappendixnumbering

\section{Sample of Neptunes}\label{app:sample}

\added{Table~\ref{tab:sample} shows the properties of the systems considered in our statistical analysis.}

\begin{longrotatetable}
\begin{deluxetable}{lcccccccc}
\tabletypesize{\scriptsize}
\tablecaption{Properties of the systems included in our sample for the statistical analysis. These parameters come from TEPCat \citep{Southworth2011} and, in particular, the obliquities from the listed references.\label{tab:sample}}
\tablewidth{0pt}
\tablehead{
\colhead{Planet} &
\colhead{$T_{\rm eff}$} &
\colhead{Period} &
\colhead{Mass} &
\colhead{Radius} &
\colhead{$\lambda$} &
\colhead{Single} &
\colhead{Compact} &
\colhead{Obliquity} \\
\colhead{} &
\colhead{(K)} &
\colhead{(days)} &
\colhead{($M_{\oplus}$)} &
\colhead{($R_{\oplus}$)} &
\colhead{($^{\circ}$)} &
\colhead{star} &
\colhead{system} &
\colhead{reference}
}
\startdata
$\pi$ Men c & $5998 \pm 62$ & $6.267852 \pm 0.000016$ & $3.623 \pm 0.381$ & $2.145 \pm 0.015$ & $-24.0 \pm 4.1$ & Y & N & \citet{Kunovac2021} \\
AU Mic b & $3540_{-110}^{+120}$ & $8.463143 \pm 0.000005$ & $8.995_{-2.670}^{+2.606}$ & $4.618 \pm 0.146$ & $-4.7_{-6.4}^{+6.8}$ & Y & Y & \citet{Hirano2020} \\
AU Mic c & $3540_{-110}^{+120}$ & $18.858827 \pm 0.000050$ & $14.461_{-3.433}^{+3.242}$ & $2.757_{-0.291}^{+0.258}$ & $68.0_{-49.0}^{+32.0}$ & Y & Y & \citet{Yu2025} \\
DS Tuc b & $5598_{-59}^{+28}$ & $8.138268 \pm 0.000011$ & $< 14.398$ & $5.604 \pm 0.224$ & $2.9_{-0.9}^{+0.9}$ & N & N & \citet{Zhou2020} \\
GJ 436 b & $3416 \pm 54$ & $2.643898 \pm 0.000000$ & $25.394_{-2.002}^{+2.098}$ & $4.102 \pm 0.157$ & $114.0_{-17.0}^{+23.0}$ & Y & N & \citet{Bourrier2022} \\
GJ 3470 b & $3725 \pm 54$ & $3.336653 \pm 0.000000$ & $13.730 \pm 1.621$ & $3.878 \pm 0.325$ & $101.0_{-14.0}^{+29.0}$ & Y & N & \citet{Stefansson2022} \\
HAT-P-11 b& $4780 \pm 50$ & $4.887802 \pm 0.000000$ & $25.013 \pm 1.526$ & $5.006 \pm 0.066$ & $133.9_{-8.3}^{+7.1}$ & Y & N & \citet{Bourrier2023} \\
HATS-38 b & $5662 \pm 80$ & $4.375040 \pm 0.000020$ & $20.659_{-4.450}^{+4.767}$ & $6.703 \pm 0.202$ & $-108.0_{-16.0}^{+11.0}$ & Y & N & \citet{Espinoza-Retamal2024} \\
HD 3167 c & $5300 \pm 73$ & $29.846495 \pm 0.000016$ & $10.679_{-0.795}^{+0.858}$ & $2.919_{-0.048}^{+0.049}$ & $-108.9_{-5.5}^{+5.4}$ & Y & Y & \citet{Bourrier2021} \\
HD 63433 b & $5700_{-75}^{+76}$ & $7.107938 \pm 0.000005$ & $2.384_{-1.812}^{+2.988}$ & $2.112_{-0.086}^{+0.093}$ & $8.0_{-45.0}^{+33.0}$ & Y & Y & \citet{Mann2020} \\
HD 93963 c & $5987 \pm 64$ & $3.645139 \pm 0.000003$ & $19.070 \pm 3.814$ & $3.230 \pm 0.072$ & $14.0_{-19.0}^{+17.0}$ & N & Y & \citet{Teng2025} \\
HD 106315 c & $6364 \pm 87$ & $21.056520 \pm 0.000120$ & $12.077 \pm 3.814$ & $4.383 \pm 0.090$ & $-2.7_{-2.6}^{+2.7}$ & Y & Y & \citet{Bourrier2023} \\
HD 110067 c & $5266 \pm 64$ & $13.673694 \pm 0.000024$ & $< 6.357$ & $2.388 \pm 0.036$ & $6.0_{-26.0}^{+24.0}$ & Y & Y & \citet{Zak2024} \\
HD 148193 b & $6369_{-69}^{+35}$ & $20.380799 \pm 0.000016$ & $38.457 \pm 9.217$ & $8.003 \pm 0.347$ & $14.0_{-8.0}^{+7.0}$ & N & Y & \citet{Knudstrup2024} \\
HD 191939 b & $5427 \pm 50$ & $8.880326 \pm 0.000005$ & $10.012 \pm 0.699$ & $3.410 \pm 0.075$ & $3.7 \pm 5.0$ & Y & Y & \citet{Lubin2024} \\
K2-25 b & $3207 \pm 58$ & $3.484565 \pm 0.000000$ & $24.473_{-5.085}^{+5.721}$ & $3.430 \pm 0.123$ & $3.0 \pm 16.0$ & Y & N & \citet{Stefansson2020} \\
K2-33 b & $3540 \pm 70$ & $5.424865 \pm 0.000033$ & $< 1175.965$ & $4.999_{-0.370}^{+0.336}$ & $-10.0_{-24.0}^{+22.0}$ & Y & N & \citet{Hirano2024} \\
K2-93 d & $6320_{-30}^{+60}$ & $278.360000 \pm 0.001500$ & $12.077 \pm 2.860$ & $9.203 \pm 0.101$ & $57.0_{-18.0}^{+26.0}$ & Y & Y & \citet{Grouffal2022} \\
K2-105 b & $5430 \pm 70$ & $8.267004 \pm 0.000022$ & $< 88.992$ & $4.136_{-0.381}^{+0.437}$ & $-81.0_{-47.0}^{+50.0}$ & Y & N & \citet{Bourrier2023} \\
K2-234 b& $5622_{-71}^{+70}$ & $11.814300 \pm 0.000200$ & $38.457_{-6.039}^{+5.721}$ & $7.263_{-0.314}^{+0.325}$ & $74.0_{-33.0}^{+34.0}$ & Y & N & \citet{Bourrier2023} \\
K2-290 b & $6302 \pm 120$ & $9.211700 \pm 0.000200$ & $< 21.104$ & $3.060 \pm 0.157$ & $173.0_{-53.0}^{+45.0}$ & N & Y & \citet{Hjorth2021} \\
Kepler-9 b & $5774 \pm 60$ & $19.225900 \pm 0.000048$ & $43.384_{-2.002}^{+1.589}$ & $8.250 \pm 0.090$ & $-13.0 \pm 16.0$ & Y & Y & \citet{Wang2018} \\
Kepler-25 c & $6270 \pm 79$ & $12.720374 \pm 0.000000$ & $15.224_{-1.621}^{+1.303}$ & $5.217_{-0.065}^{+0.069}$ & $-0.9_{-6.4}^{+7.7}$ & N & Y & \citet{Bourrier2023} \\
Kepler-30 b & $5498 \pm 54$ & $29.218700 \pm 0.000900$ & $9.185 \pm 0.095$ & $3.755 \pm 0.179$ & $4.0 \pm 10.0$ & Y & Y & \citet{Sanchis-Ojeda2012} \\
Kepler-1656 b & $5731 \pm 60$ & $31.578659 \pm 0.000007$ & $47.674_{-3.178}^{+6.357}$ & $5.022 \pm 0.527$ & $35.0_{-22.0}^{+15.0}$ & Y & N & \citet{Rubenzahl2024} \\
TOI-1136 d & $5770 \pm 50$ & $12.519370 \pm 0.000390$ & $5.594_{-0.985}^{+0.890}$ & $4.627_{-0.089}^{+0.072}$ & $5.0 \pm 5.0$ & Y & Y & \citet{Dai2023} \\
TOI-1694 b & $5135 \pm 50$ & $3.770179 \pm 0.000058$ & \nodata & $5.459_{-0.796}^{+0.471}$ & $9.0_{-18.0}^{+22.0}$ & Y & N & \citet{Handley2025} \\
TOI-1710 b& $5775 \pm 80$ & $24.283364 \pm 0.000010$ & $19.100_{-2.200}^{+2.300}$ & $5.060 \pm 0.060$ & $179.0 \pm 19.0$ & N & N & \citet{Espinoza-Retamal2026} \\
TOI-1759 b& $4065 \pm 51$ & $18.850048 \pm 0.000011$ & $10.806 \pm 1.494$ & $3.251 \pm 0.112$ & $4.0_{-18.0}^{+19.0}$ & N & N & \citet{Polanski2025} \\
TOI-181 b& $4820 \pm 80$ & $4.532054 \pm 0.000001$ & $49.900_{-6.100}^{+6.400}$ & $6.700 \pm 0.200$ & $32.0 \pm 7.0$ & Y & N &  This work\\
TOI-2076 b & $5200 \pm 100$ & $10.355230 \pm 0.000010$ & $< 8.899$ & $2.538 \pm 0.036$ & $-3.0_{-16.0}^{+15.0}$ & Y & Y & \citet{Frazier2023} \\
TOI-3884 b& $3180 \pm 88$ & $4.544584 \pm 0.000001$ & $32.736 \pm 7.310$ & $6.434 \pm 0.202$ & $37.3 \pm 1.5$ & Y & N & \citet{Chakraborty2025} \\
TOI-421 b & $5291 \pm 64$ & $5.197576 \pm 0.000005$ & $6.706 \pm 0.604$ & $2.645 \pm 0.078$ & $-35.0_{-23.0}^{+26.0}$ & N & Y & \citet{Bourrier2025} \\
TOI-421 c & $5291 \pm 64$ & $16.067541 \pm 0.000004$ & $14.112 \pm 1.398$ & $5.089 \pm 0.067$ & $14.0 \pm 1.8$ & N & Y & \citet{Bourrier2025} \\
TOI-5126 b & $6150_{-130}^{+110}$ & $5.458838 \pm 0.000007$ & \nodata & $4.741_{-0.135}^{+0.157}$ & $1.0 \pm 48.0$ & Y & Y & \citet{Radzom2024} \\
TOI-880 c & $5050 \pm 100$ & $6.387270 \pm 0.000007$ & \nodata & $4.932 \pm 0.202$ & $7.4_{-7.2}^{+6.8}$ & N & Y & \citet{Zhang2025} \\
TOI-883 b& $5697 \pm 80$ & $10.057733 \pm 0.000004$ & $39.500 \pm 2.100$ & $7.120 \pm 0.130$ & $22.0_{-14.0}^{+16.0}$ & Y & N &  This work\\
TOI-942 b & $4928_{-85}^{+125}$ & $4.324210 \pm 0.000019$ & $< 15.891$ & $3.890_{-0.235}^{+0.247}$ & $1.0_{-33.0}^{+41.0}$ & Y & Y & \citet{Wirth2021} \\
TOI-942 c & $4928_{-85}^{+125}$ & $10.156272 \pm 0.000036$ & $< 38.139$ & $4.674_{-0.303}^{+0.336}$ & $24.0 \pm 14.0$ & Y & Y & \citet{Teng2024} \\
V1298 Tau c & $5050 \pm 100$ & $8.248720 \pm 0.000024$ & $19.705_{-8.899}^{+9.217}$ & $5.235 \pm 0.235$ & $4.9_{-15.1}^{+15.0}$ & N & Y & \citet{Feinstein2021} \\
WASP-107 b& $4425 \pm 70$ & $5.721492 \pm 0.000000$ & $30.512 \pm 1.589$ & $10.357 \pm 0.247$ & $118.0_{-19.0}^{+38.0}$ & Y & N & \citet{Rubenzahl2021} \\
WASP-139 b& $5233 \pm 60$ & $5.924271 \pm 0.000001$ & $37.504 \pm 5.403$ & $8.855 \pm 0.112$ & $-85.6_{-4.2}^{+7.7}$ & N & N & \citet{Espinoza-Retamal2024} \\
WASP-156 b& $4910 \pm 61$ & $3.836160 \pm 0.000000$ & $40.682_{-2.860}^{+3.178}$ & $5.717 \pm 0.224$ & $-15.0_{-16.0}^{+17.0}$ & Y & N & \citet{Lafarga2026} \\
WASP-166 b& $6050 \pm 50$ & $5.443542 \pm 0.000003$ & $32.418 \pm 1.271$ & $7.062 \pm 0.336$ & $-15.5_{-2.8}^{+2.9}$ & Y & N & \citet{Doyle2022} \\
WASP-193 b& $6080 \pm 130$ & $6.246334 \pm 0.000000$ & $35.597_{-10.806}^{+9.217}$ & $14.785_{-0.516}^{+0.650}$ & $17.0_{-16.0}^{+17.0}$ & N & N & \citet{Yee2025} \\
\enddata
\end{deluxetable}
\end{longrotatetable}

\end{document}